\documentclass[nopreprintline]{elsarticle}

\usepackage{amsmath,amssymb,amsfonts}%

\usepackage{hyperref}
\usepackage{graphicx}
\usepackage{xcolor}

\newcommand{\lms}{\{\!\!\{}
\newcommand{\rms}{\}\!\!\}}
\newcommand{\bm}[1]{\boldsymbol{#1}}

\begin{document}

\begin{frontmatter}
\title{Mesh-Informed Reduced Order Models for Aneurysm Rupture Risk Prediction}

\author[1]{Giuseppe Alessio D'Inverno}
\author[2]{Saeid Moradizadeh}
\author[1,3]{Sajad Salavatidezfouli}
\author[1]{Pasquale Claudio Africa\corref{corresponding}}
\author[1]{Gianluigi Rozza}

\cortext[corresponding]{Corresponding author. \url{pafrica@sissa.it}}

\affiliation[1]{organization={mathLab, Mathematics Area, International School for Advanced Studies (SISSA), Trieste, Italy}}
\affiliation[2]{organization={Department of Mechanical Engineering, Islamic Azad University of Takestan, Takestan, Iran}}
\affiliation[3]{organization={Department of Electrical and Computer Engineering - Signal Processing and Machine Learning, Aarhus University, Aarhus, Denmark}}

\begin{abstract}
The complexity of the cardiovascular system needs to be accurately reproduced in order to promptly acknowledge health conditions; to this aim, advanced multifidelity and multiphysics numerical models are crucial. On one side, Full Order Models (FOMs) deliver accurate hemodynamic assessments, but their high computational demands hinder their real-time clinical application. In contrast, Reduced Order Models (ROMs) provide more efficient yet accurate solutions, essential for personalized healthcare and timely clinical decision-making. In this work, we explore the application of computational fluid dynamics (CFD) in cardiovascular medicine by integrating FOMs with ROMs for predicting the risk of aortic aneurysm growth and rupture. Wall Shear Stress (WSS) and the Oscillatory Shear Index (OSI), sampled at different growth stages of the thoracic aortic aneurysm, are predicted by means of Graph Neural Networks (GNNs). GNNs exploit the natural graph structure of the mesh obtained by the Finite Volume (FV) discretization, taking into account the spatial local information, regardless of the dimension of the input graph. Our experimental validation framework yields promising results, confirming our method as a valid alternative that overcomes the curse of dimensionality.
\end{abstract}

\begin{keyword}
thoracic aortic aneurysm \sep wall shear indices \sep computational fluid dynamics \sep reduced order models \sep graph neural networks

\MSC 65M08 \sep 68T07 \sep 92B20 \sep 76-10 \sep 76Z05 \sep 92C50

\end{keyword}
\end{frontmatter}

\section{Introduction}
\label{sec:intro}

Cardiovascular diseases (CVDs) continue to be a major global health issue, with estimations predicting that by 2030, they will account for 23.6 million deaths annually \citep{zhang2013challenges}. Aneurysms, among the most critical forms of CVDs, involve a permanent and localized expansion in the vascular wall, typically occurring in the middle part of the aorta. This expansion leads to irreversible structural changes in the vessel's wall, posing significant health risks \citep{lederle1997prevalence}.

Thoracic aortic aneurysms (TAAs) represent a severe form of aneurysm that manifests as unrepairable bulges in the thoracic aorta. These are particularly prevalent among the elderly, necessitating careful and regular monitoring of their size, especially in the case of small TAAs \citep{moll2011management}. The progression and potential rupture of aneurysms are critical concerns, and their management often involves balancing the risks of surgical intervention against the potential for catastrophic vascular failure.

Given the invasive nature of traditional experimental studies, there has been a growing interest in mathematical modeling of blood fluid dynamics as a non-invasive alternative. These models are valuable for predicting, examining, and enhancing our understanding of fluid patterns within regions affected by cardiovascular pathologies \citep{sabernaeemi2023influence, salavatidezfouli2023investigation}. Mathematical and computational approaches provide significant insights into the hemodynamics and structural mechanics of aneurysms, offering a pathway to better treatment and management strategies.

The deformation of aneurysms is a crucial aspect of their clinical management. Understanding the dynamics of aneurysm growth and deformation can significantly impact treatment decisions and patient outcomes \citep{hariri2023effects}. Previous research has employed real 3D models to study hemodynamics within aneurysms. For example, Sabernaeemi et al. \citep{sabernaeemi2023influence} and Voss et al. \citep{voss2019stent} have conducted significant studies in this area. However, these studies often present limited computational results regarding the effects of aneurysm sac growth, indicating a need for further exploration.

In this context, Wall Shear Stress (WSS) and Oscillatory Shear Index (OSI) are critical hemodynamic parameters for predicting aneurysm growth and assessing the risk of aneurysm rupture \citep{liang2023,de2020deciphering,condemi2017fluid,liang2019towards,voss2019stent,sabernaeemi2023influence}. WSS refers to the tangential force exerted by blood flow on the vessel wall, influencing endothelial cell function and vascular remodeling. Abnormal WSS can contribute to the development and progression of aneurysms by promoting inflammatory responses and weakening the vessel wall structure. OSI, on the other hand, measures the change in the direction of shear stress over the cardiac cycle, indicating areas of disturbed flow that are prone to pathological changes. High OSI values are often associated with regions of the arterial wall that are susceptible to aneurysm formation and growth due to prolonged exposure to oscillatory and low shear stresses. Understanding the spatial and temporal variations of WSS and OSI within aneurysms is essential for identifying high-risk regions and improving the prediction of aneurysm growth and rupture, thereby aiding in developing targeted therapeutic interventions and enhancing patient outcomes.

Traditionally, computational fluid dynamics (CFD) for cardiovascular modeling has relied on methods such as the finite element method (FEM), finite volume method (FV), and finite difference method (FD). These techniques have been instrumental in providing detailed simulations of fluid flow and structural interactions within the cardiovascular system. However, they often require significant computational resources and can be time-consuming, especially for complex geometries and time-dependent analyses \cite{africa2024lifex,bucelli2023,zingaro2022}.

Recent advancements in computational modeling, particularly the application of Graph Neural Networks (GNNs), offer promising new avenues for cardiovascular blood flow modeling. GNNs provide an innovative approach by leveraging the inherent graph structure of a computational mesh, allowing for efficient and accurate representations of fields defined at mesh vertices, and consequently, for predictions of hemodynamic parameters.

For instance, Pegolotti et al. \citep{pegolotti2024learning} utilized MeshGraphNets to predict pressure and flow rates at vessel centerline nodes within a 1D model. Similarly, Suk et al. \citep{suk2021equivariant} employed equivariant GNNs on CFD-derived meshes to predict WSS on the surface of single arteries. Their methodology, however, had limitations: the dataset splitting over training and test sets does not reflect any physics- or geometry-based criterion, they did not predict WSS on completely unseen arteries, did not evaluate WSS over time, and did not focus on aneurysms specifically. 
Furthermore, they primarily focused on typical machine learning metrics to evaluate the quality of the model training, rather than assessing clinically relevant indices such as the OSI.
Finally, Suk et al. \citep{suk2023se} used GNNs as an efficient black-box surrogate model to estimate 3D velocity fields in the artery lumen.

The present study aims to address these gaps by applying advanced GNN methodologies to predict hemodynamic parameters in aneurysms, with a particular focus on the deformation and growth of the aneurysm sac. By leveraging these computational tools, we aim to enhance the understanding and treatment of this critical cardiovascular condition. Specifically, this study will focus on:

\begin{enumerate}
    \item Developing and validating GNN-based models to predict wall shear indices such as WSS and OSI in aneurysms.
    \item Examining the temporal dynamics of WSS in aneurysms.
    \item Investigating the effects of aneurysm sac growth on hemodynamic parameters.
\end{enumerate}

Through these objectives, we seek to contribute to the field of cardiovascular modeling by providing robust, non-invasive tools that can aid in the diagnosis, monitoring, and treatment of aneurysms, ultimately improving patient outcomes and advancing the state of cardiovascular research. 

\section{Materials and methods}
\label{sec:methods}

\subsection{FOM}
This section introduces the mathematical model used to describe blood flow in the toracic aorta, specifically the incompressible Navier-Stokes equations. This includes adopted boundary conditions and discretization methods.

\subsubsection{Mathematical Model}\label{subsec:math_model}
The blood motion is determined within a time-independent domain \( \Omega \) over a time interval \([0, T]\). We assume, without loss of generality, that the domain boundary $\partial \Omega$ is partitioned into 4 subsets as $\partial \Omega = \Gamma^{\mathrm{in}} \cup \Gamma^{\mathrm{out}}_{\mathrm{SA}} \cup \Gamma^{\mathrm{out}}_{\mathrm{abd}} \cup \Gamma^{\mathrm{wall}}$, as shown in Figure \ref{FIG_Mesh}. The flow is governed by the following equations:
\begin{equation}
\label{eq_NS}
\begin{array}{ll}
\rho \partial_t \bm{u}+\rho \nabla \cdot(\bm{u} \otimes \bm{u})-\nabla \cdot \bm{\sigma}=0 & \text { in } \Omega \times \left(0, T\right],
\end{array}
\end{equation}
\begin{equation}
\label{eq_cont}
\begin{array}{ll}
\nabla \cdot \bm{u}=0 & \text { in } \Omega \times \left(0, T\right], 
\end{array}
\end{equation}
where $\rho=1060 \, \text{kg/m}^3$ is the blood density, \(\mathbf{u}\) represents the velocity vector, and \(\bm{\sigma}\) denotes the Cauchy stress tensor. Eq. \ref{eq_NS} expresses the conservation of momentum, while Eq. \ref{eq_cont} ensures the mass conservation. The blood is modeled as a non-Newtonian fluid, where \(\bm{\sigma}\) is defined as:
\begin{equation}
\label{eq_cauchy}
\begin{array}{ll}
\bm{\sigma}(\bm{u}, p)=-p\mathbf{I} +\bm{\tau}, 
\end{array}
\end{equation}
with $\bm{\tau}$ representing the shear stress tensor:
\begin{equation}
\label{stress_tensor}
\begin{array}{ll}
\bm{\tau}=2\mu(\dot{\gamma})\mathbf{D}, 
\end{array}
\end{equation}
where $\mathbf{D} = \frac{1}{2} \left(\nabla \mathbf{u}+\nabla \mathbf{u}^{\mathrm{T}}\right)$, \(p\) represents the pressure, $\mu$ is the dynamic viscosity, and $\dot{\gamma}$ is the shear rate:
\begin{equation}
\label{eq_shearRate}
\begin{aligned}
\dot{\gamma} & =\sqrt{\frac{1}{2} \mathbf{D}: \mathbf{D}}.
\end{aligned}
\end{equation}

For a Newtonian fluid, $\mu$ is a constant and independent of the shear rate, while for a non-Newtonian fluid, it is a function of $\dot{\gamma}$. The viscosity of blood is modeled using the non-Newtonian Casson model:
\begin{equation}
\label{eq_Casson}
\mu=\frac{\bm{\tau}_0}{\dot{\gamma}} +\frac{\sqrt{\mu_{\infty} \bm{\tau}_0}}{\sqrt{\dot{\gamma}}}+\mu_{\infty},
\end{equation}
where $\bm{\tau}_0$ and $\mu_{\infty}$ are 0.005 and 0.0035, respectively. Accounting for the non-Newtonian rheology of blood enables a more precise characterization of the hemodynamic indicators described below in the thoracic aorta, particularly under pathological conditions, as demonstrated in \cite{petuchova2022comparison,febina2018wall}.

To investigate blood flow patterns, we define the wall shear stress as \citep{liang2023}:
\begin{equation}
\label{eq_WSS}
\begin{array}{ll}
\mathbf{WSS}=\bm{\tau}_w \cdot \mathbf{n},
\end{array}
\end{equation}
where \(\mathbf{n}\) is the normal vector, and $\bm{\tau}_w := \bm{\tau}_{|\Gamma^{\mathrm{wall}}}$ denotes the wall shear stress tensor. 

A scalar quantity of interest is represented by the so-called Time-Averaged Wall Shear Stress (TAWSS), computed as the averaged integral of the WSS over time:

\begin{equation}\label{eq:TAWSS}
    \text{TAWSS} = \frac{1}{T} \int_0^T |\mathbf{WSS}|dt
\end{equation}

The OSI quantifies the cyclic deviation of the Wall Shear Stress vector from its predominant direction. 
In the time-dependent formulation, the OSI is defined as:

\begin{equation}\label{eq:OSI_t}
\begin{gathered}
\text { OSI}(t)=\frac{1}{2}\left(1-\frac{\left|\frac{1}{T} \int_{t-T}^T \mathbf{WSS}\,dt\right|}{\frac{1}{T} \int_{t-T}^T|\mathbf{WSS}| \, dt}\right).
\end{gathered}
\end{equation}

In literature, OSI is often defined evaluating Equation \ref{eq:OSI_t} in the final timestamp $T$:

\begin{equation}\label{eq:OSI}
\begin{gathered}
\text { OSI }=\frac{1}{2}\left(1-\frac{\left|\frac{1}{T} \int_{0}^T \mathbf{WSS}\,dt\right|}{\frac{1}{T} \int_{0}^T|\mathbf{WSS}| \, dt}\right).
\end{gathered}
\end{equation}

The index ranges from 0 to 0.5 where the value of 0 indicates the perfect alignment of the instantaneous vector with the time-averaged vector, while the value of 0.5 means that the instantaneous vector is never aligned with the time-averaged vector. Put differently, OSI measures flow reversal and is insensitive to the shear magnitude.

For the sake of simplicity, in what follows we denote by $\text{WSS}$ the magnitude of $\mathbf{WSS}$.

\subsubsection{Turbulence}
The flow in the TAA can be considered laminar under resting conditions, with a transition to turbulence occurring during peak systole \citep{etli2021numerical}. When modeling the flow in TAA, it is crucial to consider that pure laminar or fully turbulent models, such as the $k-\omega$ or $k-\epsilon$ models, may not yield accurate results \citep{kelly2020influence}. The transitional $k-\omega$ SST model extends the well-known SST (Shear Stress Transport) turbulence model by incorporating two additional equations: intermittency, $\gamma$, and the transition momentum thickness Reynolds number, $\text{Re}_{\theta t}$. These equations capture the physics of the transition, i.e. laminar-turbulent and vice-versa. This transitional $k-\omega$ SST model has been successfully applied in numerous cardiovascular studies to simulate blood flow within the aorta \citep{zhu2022fluid, sengupta2023aortic, salavatidezfouli2024effect}. Therefore, this model was chosen for all simulations. Further details on the model's formulation are available in \citep{johari2019disturbed}.

\subsubsection{Numerical discretization}
To solve the blood flow in the 3D aortic model, ANSYS FLUENT, a commercial package based on the cell-centered finite volume method, has been employed. Finite volume is preferred for 3D complex models of fluid flow problems due to its efficiency in computational time and memory \citep{lopes2021analysis}. Both the spatial and temporal terms in Eq. \ref{eq_NS} are discretized using a second-order scheme, and the SIMPLE (Semi-Implicit Method for Pressure-Linked Equations) algorithm has been incorporated for pressure-velocity coupling. A time step of $10^{-4}$sec has been used for all simulations.

\subsection{3D Model}
\begin{figure}
\centering
\includegraphics[scale=1.1]{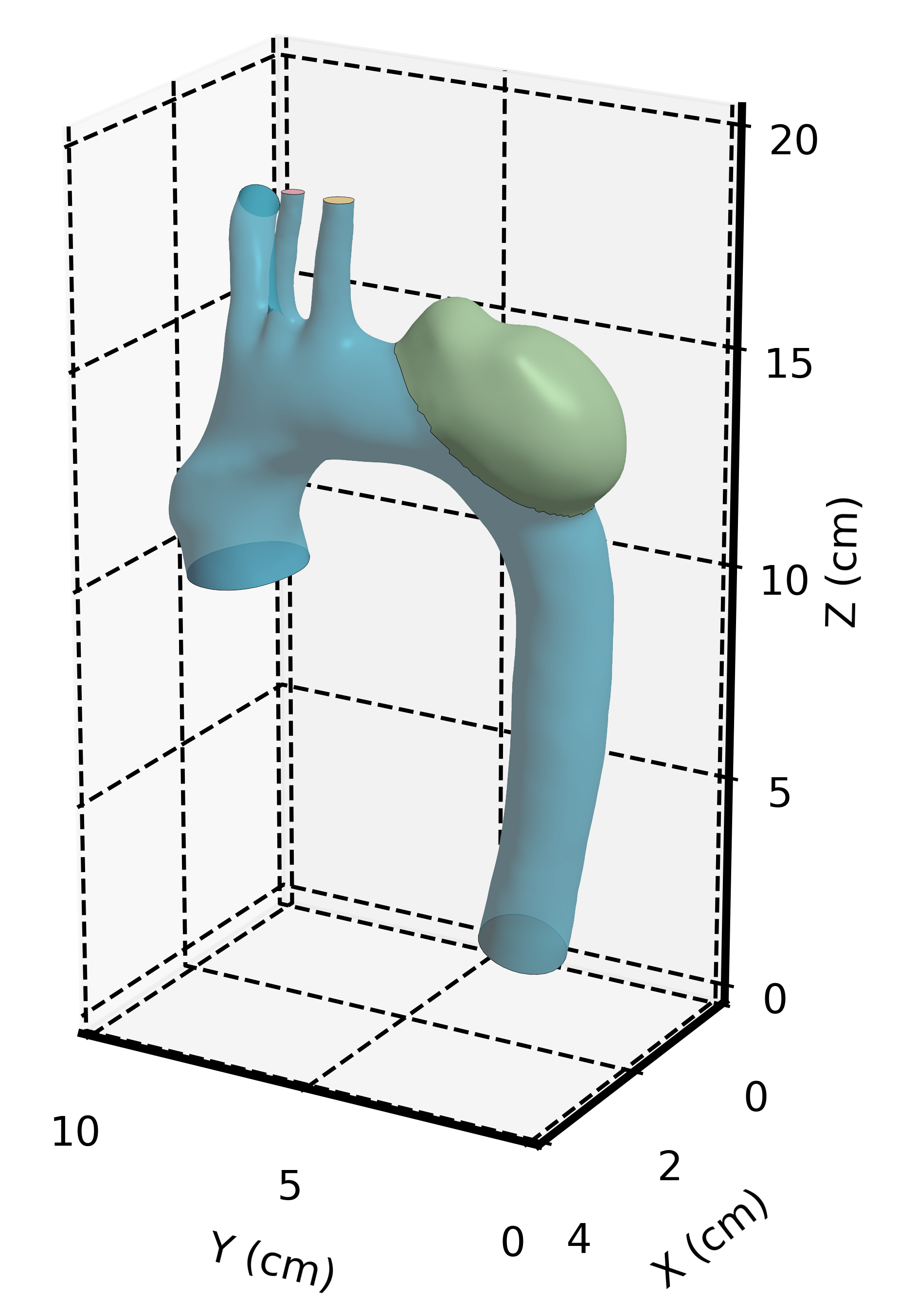}
\caption{Final geometry of TAA after clean-up.}
\label{FIG_Geom}
\end{figure}

It is well-known that the flow characteristics in the TAA strongly depend on its shape, which directly influences wall shear indices \cite{friedman1983arterial,morbiducci2016atherosclerosis}. Consequently, using patient-specific geometry is essential to obtain accurate results. To achieve this, CT-scan images of a 48-year-old male patient were acquired. Image-processing techniques were then employed to reconstruct a 3D model from the 2D images. The model is adopted from \citep{faraji2023numerical}, where all details regarding the utilized software and the image processing of the CT data are provided. Additional branches and extraneous details were removed during this process. Moreover, the outlet boundaries were extended by a specified length in the normal direction to minimize the impact of the outlet boundary condition on the main flow field and enhance numerical stability \citep{joly2018flow, su2020generating}. The final geometry of the TAA, after these modifications and clean-up, is shown in Fig. \ref{FIG_Geom}. 

\begin{figure}
\centering
\includegraphics[scale=0.08]{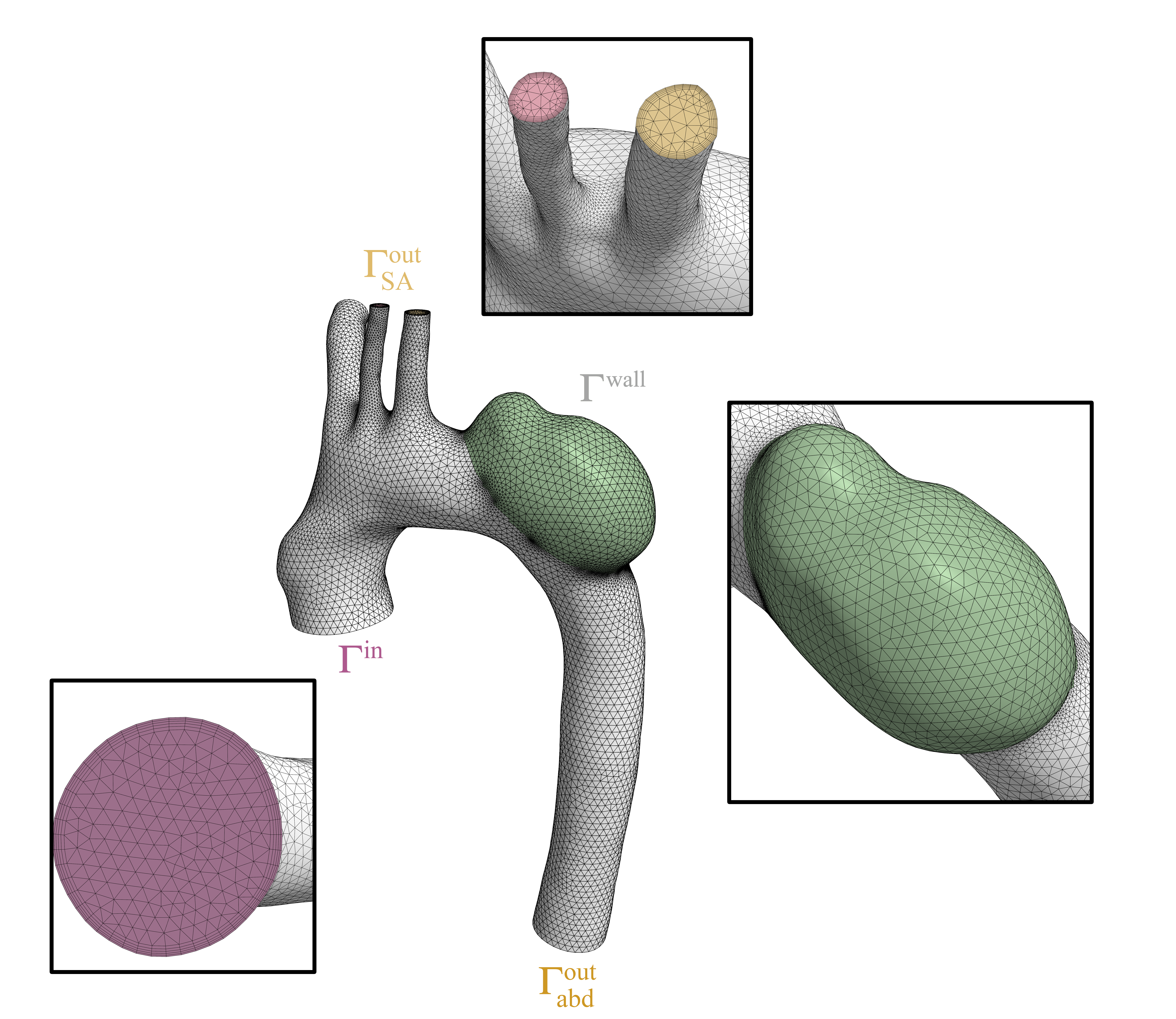}
\caption{Computational Mesh.}
\label{FIG_Mesh}
\end{figure}

Careful considerations have been taken during mesh generation. Tetrahedral elements have been created for the entire domain, with five prism layers added to the wall boundaries to capture the boundary layer effects. The computational mesh is shown in Fig. \ref{FIG_Mesh}.

\begin{figure}
\centering
\includegraphics[scale=0.5]{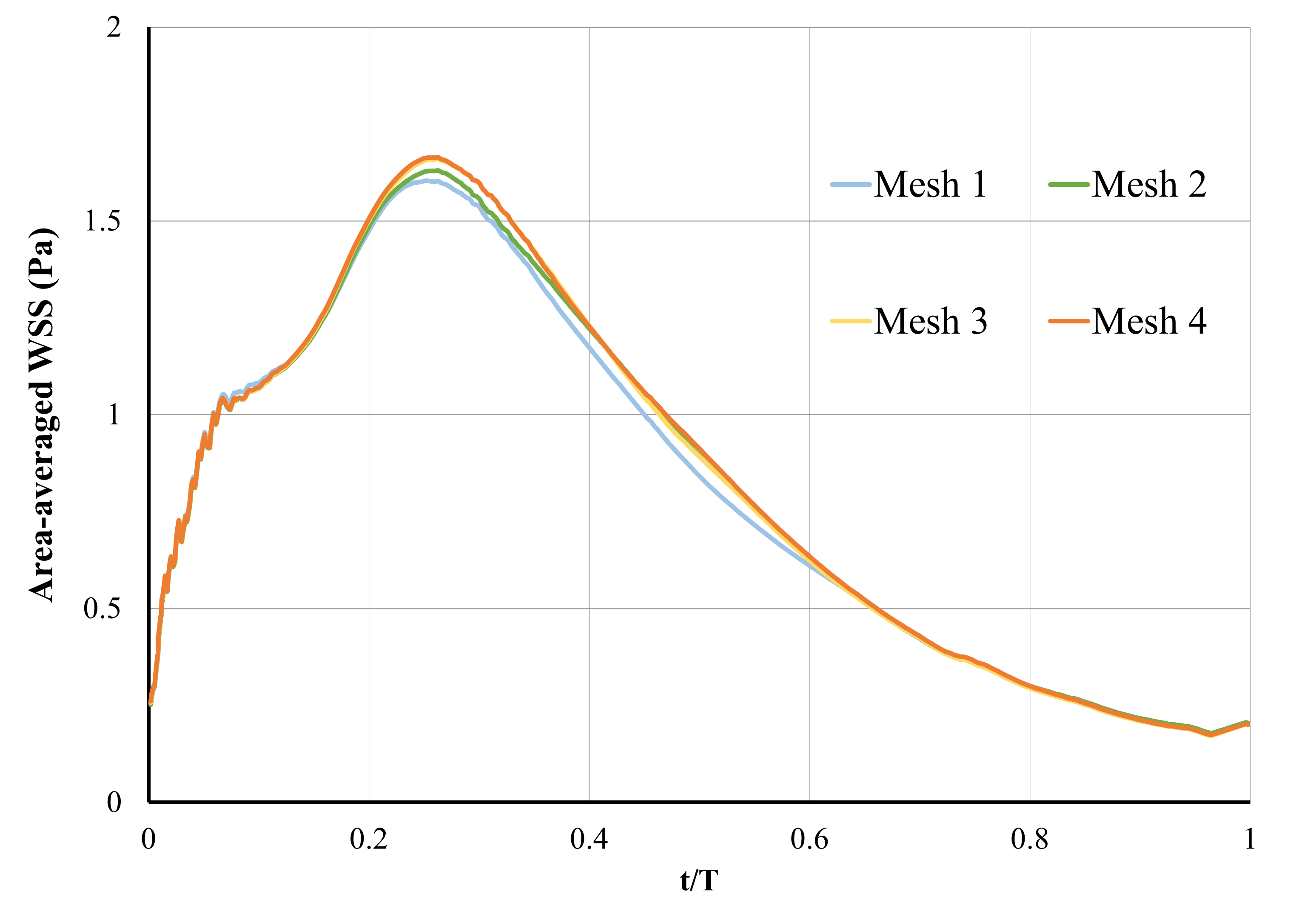}
\caption{Temporal evolution of area-averaged WSS on the aneurysm wall for different meshes.}
\label{FIG_MI}
\end{figure}

To evaluate the independence of CFD results from mesh size, four sets of meshes containing $1.2 \times 10^5$, $2.4 \times 10^5$, $3.3 \times 10^5$ and $4.5 \times 10^5$ tetrahedral elements were created. These meshes correspond to varying element sizes on the aneurysm wall, specifically $0.3$ mm, $0.2$ mm, $0.15$ mm, and $0.125$ mm, respectively. Temporal variations of the area-averaged WSS on the aneurysm wall for all mesh sizes are presented in Fig. \ref{FIG_MI}. This figure shows minimal differences between the two finest meshes, leading to the selection of the third mesh for subsequent simulations.

\begin{figure}
\centering
\includegraphics[scale=0.1]{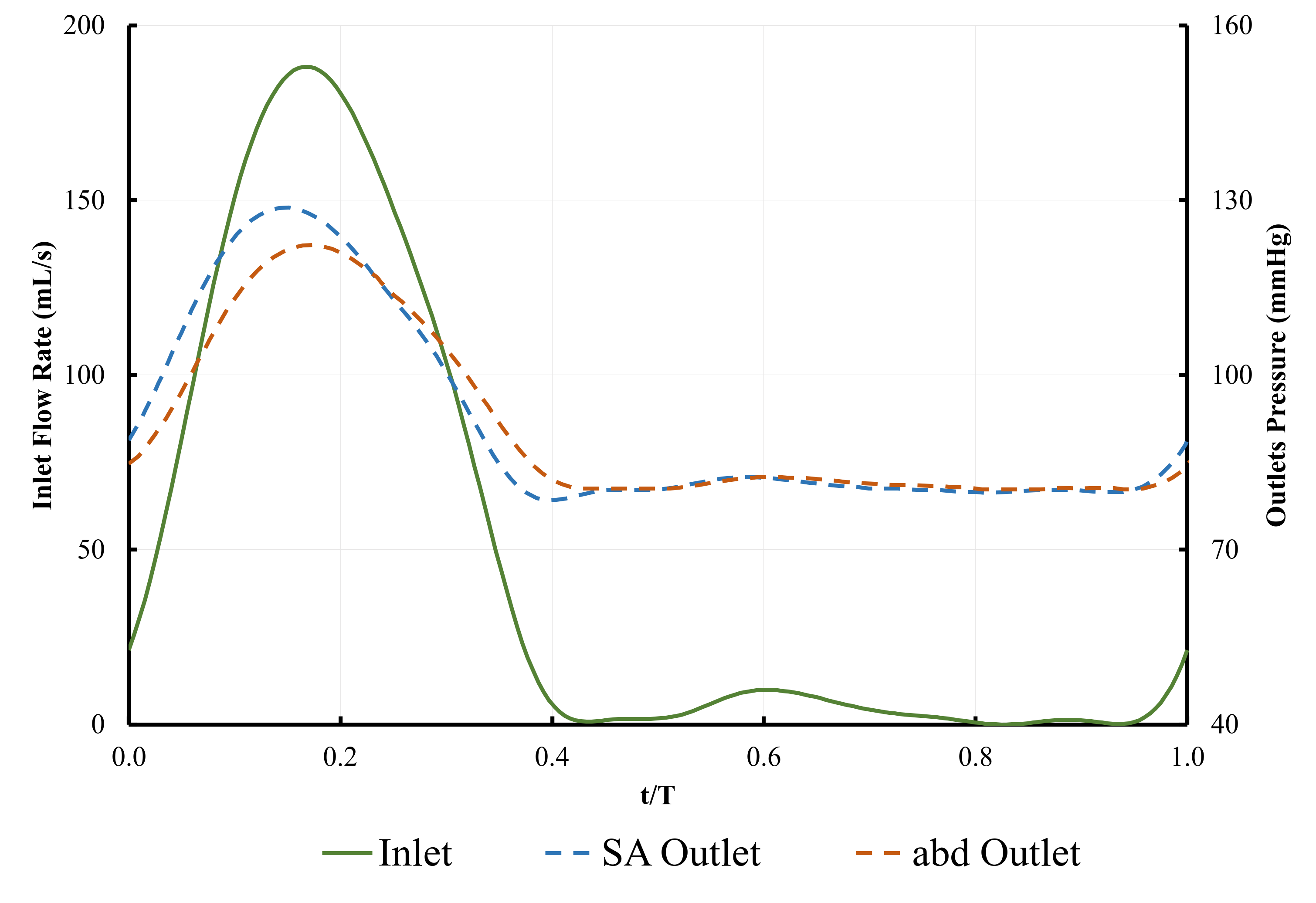}
\caption{Temporal profiles for the inlet and outlet boundary conditions.}
\label{FIG_BC}
\end{figure}

The inflow and outflow boundaries are managed by mass flow inlet and pressure outlet boundary conditions, respectively, based on data provided by \cite{africa2024lifex}. The time-dependent profile for the inlet and both top and bottom outlets is illustrated in Fig. \ref{FIG_BC}. Furthermore, only the fluid flow is modeled; thus, there is no vascular structure modeling, and the walls are assumed to be rigid. Additionally, a no-slip condition is applied on the vascular walls.

Three cardiac cycles are required to achieve fully periodic flow. Additionally, to ensure convergence, initial values of zero for pressure and velocity were applied, making the early cycles inaccurate. 

\subsection{Generation of Scaled Aneurysm Models}
\begin{figure}
\centering
\includegraphics[scale=0.15]{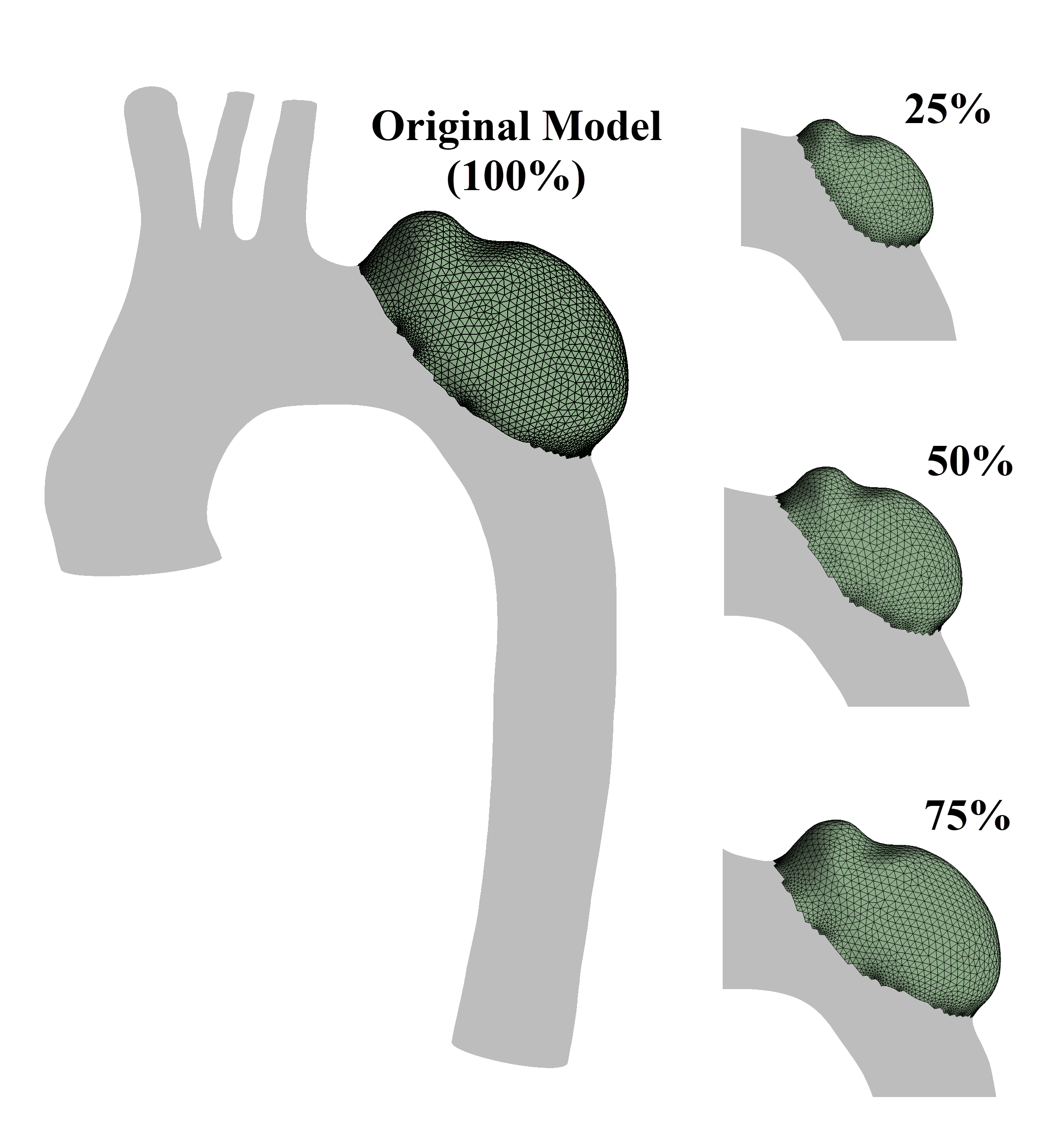}
\caption{Scaled volumes.}
\label{FIG_Volumes}
\end{figure}

The primary objective of this study is to predict wall shear indices on the aneurysm wall over an extended period as the aneurysm progresses. To achieve this, we consider a model of an aneurysm at its original volume, referred to as 100\% or the original volume. From this baseline, we artificially scale down the aneurysm volume to generate smaller sizes of 25\%, 50\%, and 75\% of the original. These scaled models, depicted in Fig. \ref{FIG_Volumes}, are intended to represent the possible growth stages of a patient’s aneurysm over different years. By evaluating these different volumes, we aim to simulate the progression of aneurysm enlargement and assess the corresponding changes in the hemodynamics properties of the aneurysm wall.

The decision to artificially generate different aneurysm volumes, rather than using patient-specific data, stems from several practical considerations. Firstly, the requisite longitudinal data capturing the gradual growth of aneurysms in individual patients over time is often not available. Furthermore, for smaller aneurysm sizes, such as those comparable to the 25\% volume, medical intervention is generally not pursued unless there are specific indications. Consequently, CT scans and other diagnostic images that could provide such data are rarely available for these preliminary sizes.

\subsection{Snapshot collection}
For each aneurysm volume size, a CFD simulation has been conducted to collect necessary data for various aneurysm sizes. 

To systematically capture the temporal dynamics of  WSS index, across the aneurysm wall, data snapshots are collected at consistent intervals throughout the CFD simulation. Specifically, snapshots are taken every five iterations for the final cycle to track changes in the variables of interest over time. 

Each snapshot comprises a vector, $\tilde{WSS}_s$, that encapsulates the local WSS values with $\mathcal{N}$ entries ($\mathcal{N}$ being the total number of nodes of the aneurysm wall boundary). These vectors are subsequently assembled into a comprehensive snapshot matrix for WSS, denoted as $\tilde{\mathbf{WSS}}$, and structured as follows:

\begin{equation}
\label{eq_WSS_mat}
\begin{array}{ll}
\tilde{\mathbf{WSS}} = \left(\tilde{WSS}_1, \tilde{WSS}_2, \ldots, \tilde{WSS}_s, \ldots, \tilde{WSS}_S\right),
\end{array}
\end{equation}
where $S$ signifies the total number of snapshots collected throughout the simulation period equal to 500. A similar process is employed to construct the $\tilde{\mathbf{OSI}}$ snapshot matrix. 

\subsection{Graph Neural Networks} \label{subsec:gnn}
Graph Neural Networks (GNNs) \cite{sperduti1997supervised} \cite{scarselli2008graph} \cite{micheli2009neural} are a widely popular class of machine learning algorithms designed to operate on structured data, such as social networks, databases, traffic networks, and molecules, suitable to be processed with a graph structure, i.e. as nodes (or vertices) connected by edges that represent the relationship between linked nodes. They have shown outstanding performances in many applications such as traffic flow predictions \cite{jiang2022graph} , molecule property prediction \cite{wieder2020compact} and generation \cite{bongini2021molecular}, and weather forecasting \cite{lam2022graphcast}. 

Most of the literature around learning over structured data concerns a particular class of GNNs, the so-called \textit{message-passing} GNNs, which many of the popular models in this field belong to, such as Graph Convolutional Networks (GCN) \cite{kipf2016semi}, GraphSAGE \cite{hamilton2017inductive}, Graph Isomorphism Networks (GIN) \cite{xu2018powerful} and Graph Attention Networks (GAT) \cite{velivckovic2017graph}. 
Formally, a message passing GNN acting on a graph domain $\mathcal{G}= \{G_1, \dots, G_n \}$ with $G_i = (V_i, E_i)$,  relies on the following recursive updating scheme:

\begin{equation}\label{def:gnn_upd}
    \mathbf{h}_v^{(t+1)} = \mathsf{COMBINE}^{(t+1)}\big(  \mathbf{h}_v^{(t)}, \mathsf{AGGREGATE}^{(t+1)} (\lms \mathbf{h}_u^{(t)} | u \in \mathsf{ne}(v) \rms)   \big), 
\end{equation}

for all $v\in V$ and $t =0 , \dots L-1 $, where $\mathbf{h}_v^{(t)} \in \mathbb{R}^d$ is the hidden feature of node $v$ at time $t$, $L$ is the number of layers of the GNN, $\lms \cdot \rms $ denotes a multiset and $\mathsf{ne}(v) = \{ u \in V \; s.t. \; (u,v) \in E \}$ .
Here $\{ \mathsf{COMBINE}^{(t)} \}_{t=1,\dots, L}$ and $\{ \mathsf{AGGREGATE}^{(t)} \}_{t=1,\dots, L}$ are functions that can be defined by learning models; intuitively, $\mathsf{AGGREGATE}^{(t+1)}$ aggregates the information from the neighborhood of a node $v$ into a hidden vector $h_{\mathsf{ne}}^{(t+1)}$, which is then combined through $ \mathsf{COMBINE}^{(t+1)}$ with the feature $h_v^{(t)}$ of node $v$ at previous iteration. Eventually, a $\mathsf{READOUT}$ function can be provided to output the prediction of the task; this function will take in input the features of all the nodes, i.e. $o=\mathsf{READOUT}(\lms \mathbf{h}_u^{(L)}| u\in V\rms )$, in case of graph-focused tasks; 
conversely, in node--focused tasks, is calculated on a specific node, i.e., $o_v=\mathsf{READOUT}( \mathbf{h}_v^{(L)} ), \, \forall v \in V$. This formulation ensures GNNs to be permutation equivariant.

GNNs have been tightly linked with the \textit{1-order Weisfeiler-Lehman} isomorphism test (1-WL), as their respective expressive powers have been proven to be equivalent \cite{xu2018powerful} \cite{morris2019weisfeiler}.

Moreover, GNNs have been proved to be universal approximators under mild assumptions \cite{d2024approximation} \cite{azizian2020expressive}, even on different domain of graphs, i.e. edge-attributed graphs and dynamic graphs \cite{beddar2024weisfeiler}; these results provide guarantees for their effective implementation in many applicative fields.

\section{Results and discussion}
\label{sec:results}

\subsection{Experimental setup}\label{subsec:setup}
In this section we describe the experimental framework dedicated to the prediction of the WSS index introduced in Section \ref{subsec:math_model}, via GNNs. Experiments were run on an 11th Gen Intel(R) Core(TM) i7-11700 running at 2.50GHz using 31 GB of RAM and a NVIDIA Quadro RTX 4000. Code is available at \url{https://github.com/AleDinve/aneurysm_gnn/}.

\paragraph*{Dataset description}

Each sample of our dataset is generated as follows:

\begin{itemize}
    \item \textit{Connectivity}: The connectivity of the surface mesh of the TAA is extracted from the Finite Volume Discretization, leading to the reference graph. The number of nodes and the connectivity depends on the specific size; values are reported in Table \ref{tab:summary_graphs} ;
    \item \textit{Node features}: for each timestamp and for each TAA size, we label the graph nodes $v$ with a feature $\mathbf{x}_v = [x,y,z, t, u(t), p_{\text{abd}}(t), p_{\text{SA}}(t)]$, where: $x,y,z$ are the coordinates of each node (that varies according to the TAA size), $t$ is the times step, $u(t)$ is the inlet flow rate, $p_{\text{abd}}(t), p_{\text{SA}}(t)$ are the abdominal and SA outlet pressure.
    \item \textit{Target features}: for each timestamp and for each TAA size,the node target features $y_v$ consist in the specific (normalized) wall shear index (WSS or OSI).
\end{itemize}

\begin{table}[!ht]
    \centering
        \caption{Summary of number of nodes and number of elements on the obtained graph at each size of the TAA.}
    \begin{tabular}{|c || c | c|}
    \hline 
        \textbf{Size} & \textbf{\# nodes} & \textbf{\# elements} \\
        \hline
       \textbf{25\%}  & 1507 & 2797  \\
        \textbf{50\%} & 1899 & 3519 \\
        \textbf{75\%} & 2177 & 4031 \\
        \textbf{100\%}  & 2859 & 5313\\   
        \hline
    \end{tabular}

    \label{tab:summary_graphs}
\end{table}

In summary, for each growth stage, we have 500 graphs (one for each time step $t)$ with node features $\mathbf{x}_v \in \mathbb{R}^7$ and node target features $y_v \in [0,1]$. 

We aim at predicting the WSS and the OSI indices over the surface of a TAA at a specific growth stage by means of GNNs. Our numerical study is split in two steps: 
\begin{itemize}
    \item the first step consists of  training a GNN over a training set composed by graph snapshots taken from TAAs of sizes 25\%,50\% and 100\%, while it is evaluated over a test set composed by graph snapshots taken from a TAA of size 75\%. This task focuses on the \textit{interpolation capabilities} of the model over shear indices at different growth stages of the TAA.
    \item the second step consists of training a GNN over a training set composed by graph snapshots taken from TAAs of sizes 25\%,50\% and 75\%, while it is evaluated over a test set composed by graph snapshots taken from a TAA of size 100\%. This task focuses on the \textit{extrapolation capabilities} of the model over shear indices at different growth stages of the TAA. This task is clinically more relevant as we would like to predict the WSS on future (and therefore larger) volume sizes of the TAA.
\end{itemize}

We choose the Mean Square Error as loss functional, and the Adam algorithm as optimizer, with the learning rate set to $\lambda = 10^{-3}$. 

\paragraph*{Choice of the GNN module}
We conducted a preliminar experimental study to determine how to choose the best GNN architecture to be used in the main experimental framework. Among the several GNN modules proposed in literature, we focused on 4 particular modules: GraphConv, Graph Convolutional Networks (GCNs), Graph Isomorphism Networks (GINs) and Graph Transformers. A detailed description of each module can be found in  
\ref{app:gnn_modules}.

In order to choose the optimal GNN module, we conduct both a quantitative and a qualitative comparison on the interpolation task. For the quantitative comparison, we compute the relative $\infty$-norm error between the TAWSS computed over the reference WSS values and the TAWSS computed over the predicted WSS values (denoted as $\overline{\text{TAWSS}}$):

\begin{equation}\label{eq:error_TAWSS}
    \text{err}_{\text{TAWSS}} = \frac{\|  \overline{\text{TAWSS}} - \text{TAWSS}\|_{\infty}}{\| \text{TAWSS} \|_{\infty}}
\end{equation}

Statistics of the obtained $\text{err}_{\text{TAWSS}}$ over 5 seeds are reported in Table \ref{tab:TAWSS_architecture}.

\begin{table}[!ht]
    \centering
        \caption{Main statistics of $\text{err}_{\text{TAWSS}}$ over 5 runs, for each employed GNN module. }
    \begin{tabular}{|c||c|c|c|c|}
    \hline
         & \textbf{mean} & \textbf{std} & \textbf{min} & \textbf{max} \\
       \hline
        \textbf{GraphConv} & 0.427356 & 0.024287 & 0.398182 & 0.462096 \\
        \textbf{GIN} & 0.524057 & 0.029917 & 0.483391
        & 0.557245 \\
        \textbf{GCN} & 0.402119 &  0.071388 & 0.314738 & 0.510215 \\
        \textbf{Graph Transformer}  & 0.422558  & 0.042028 & 0.352242  &  0.456901 
        \\ \hline
    \end{tabular}
    \label{tab:TAWSS_architecture}
\end{table}

As it emerges from our preliminar analysis, GCN obtains the minimum mean error over the TAWSS index among all the examined GNN modules, but it also has the greatest variance, while Graph Transformer and GraphConv obtain comparable mean errors but with lower standard deviation. 

\begin{figure}[!ht]
\centering\includegraphics[width=\linewidth]{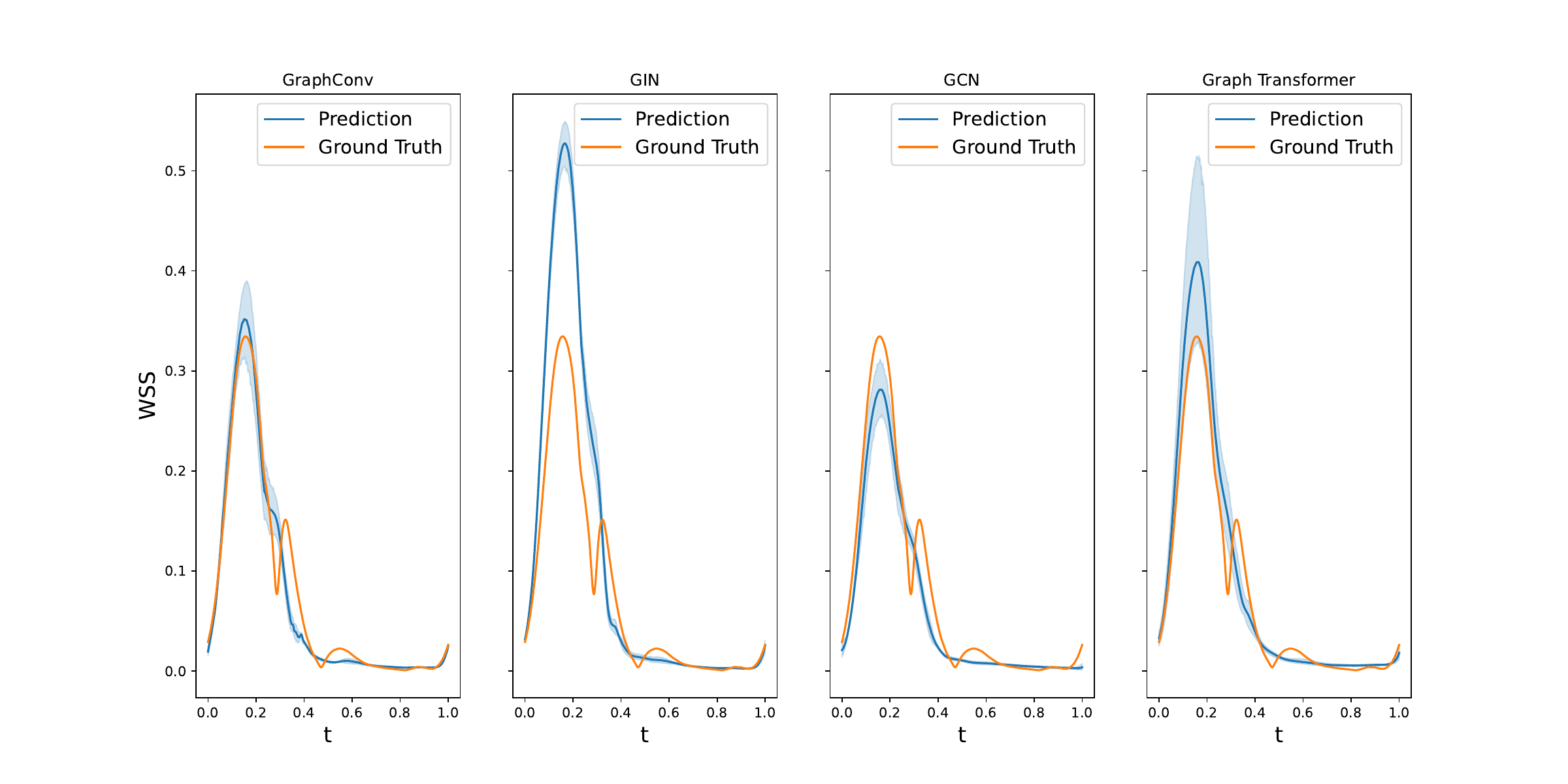}
    \caption{Prediction of the WSS index in time by different GNN modules on node 60.}
    \label{fig:node_60}
\end{figure}

\begin{figure}[!ht]
\centering\includegraphics[width=\linewidth]{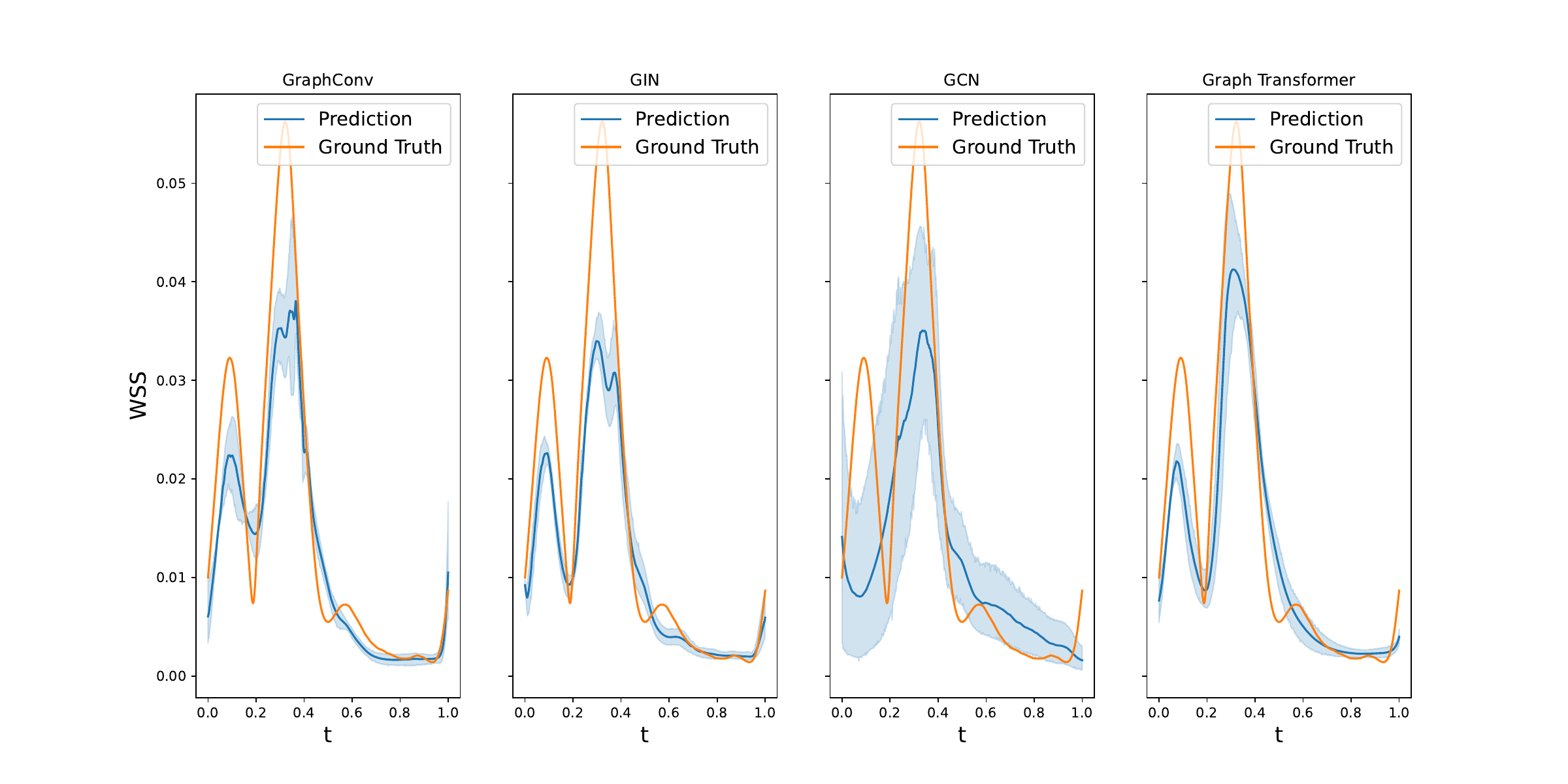}
    \caption{Prediction of the WSS index in time by different GNN modules on node 735.}
    \label{fig:node_735}
\end{figure}

\begin{figure}[!ht]
\centering\includegraphics[width=\linewidth]{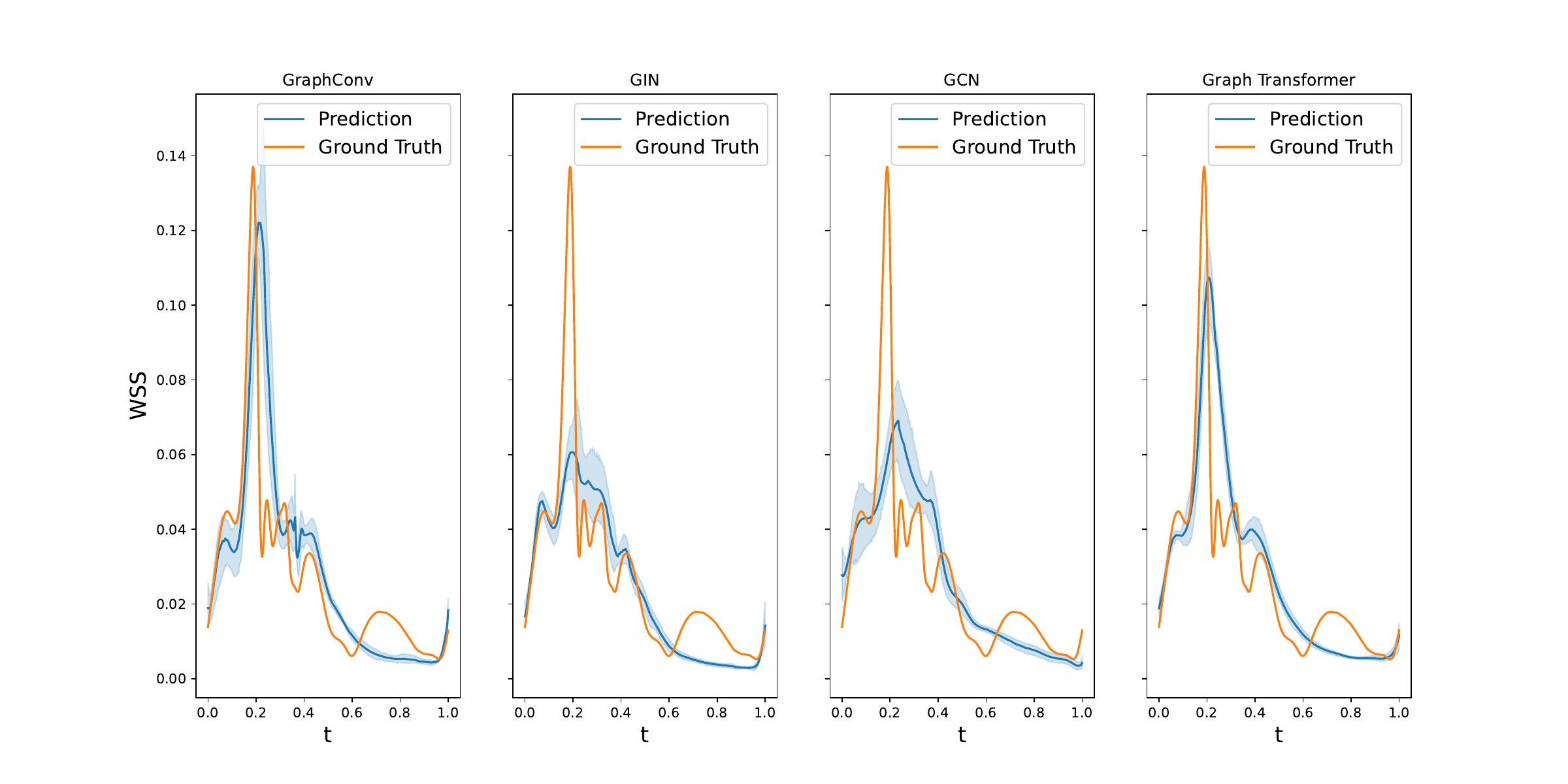}
    \caption{Prediction of the WSS index in time by different GNN modules on node 1083.}
    \label{fig:node_1083}
\end{figure}

\begin{figure}[!ht]
\centering\includegraphics[width=\linewidth]{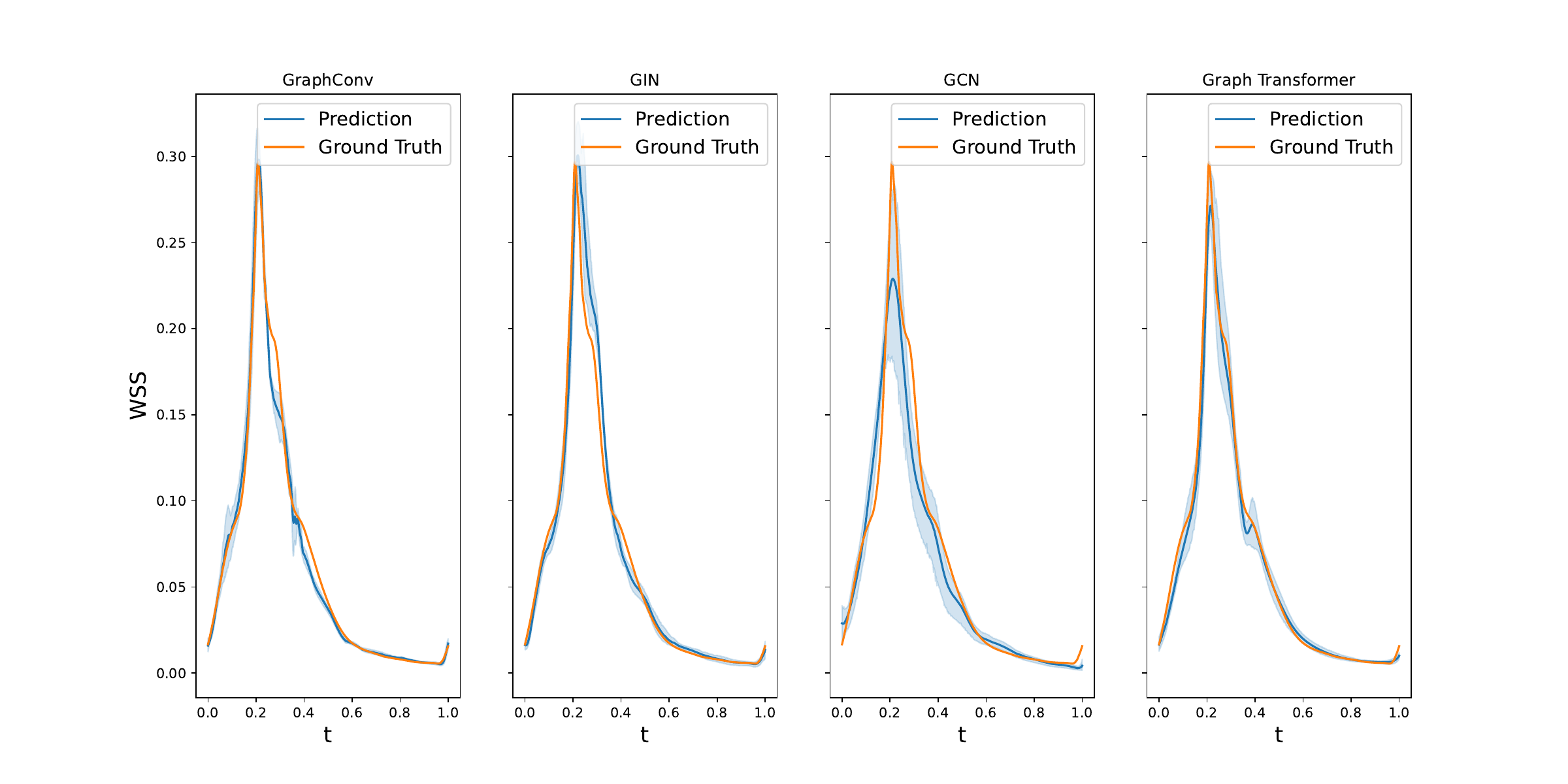}
    \caption{Prediction of the WSS index in time by different GNN modules on node 1167.}
    \label{fig:node_1167}
\end{figure}

\begin{figure}[!ht]
\centering\includegraphics[width=\linewidth]{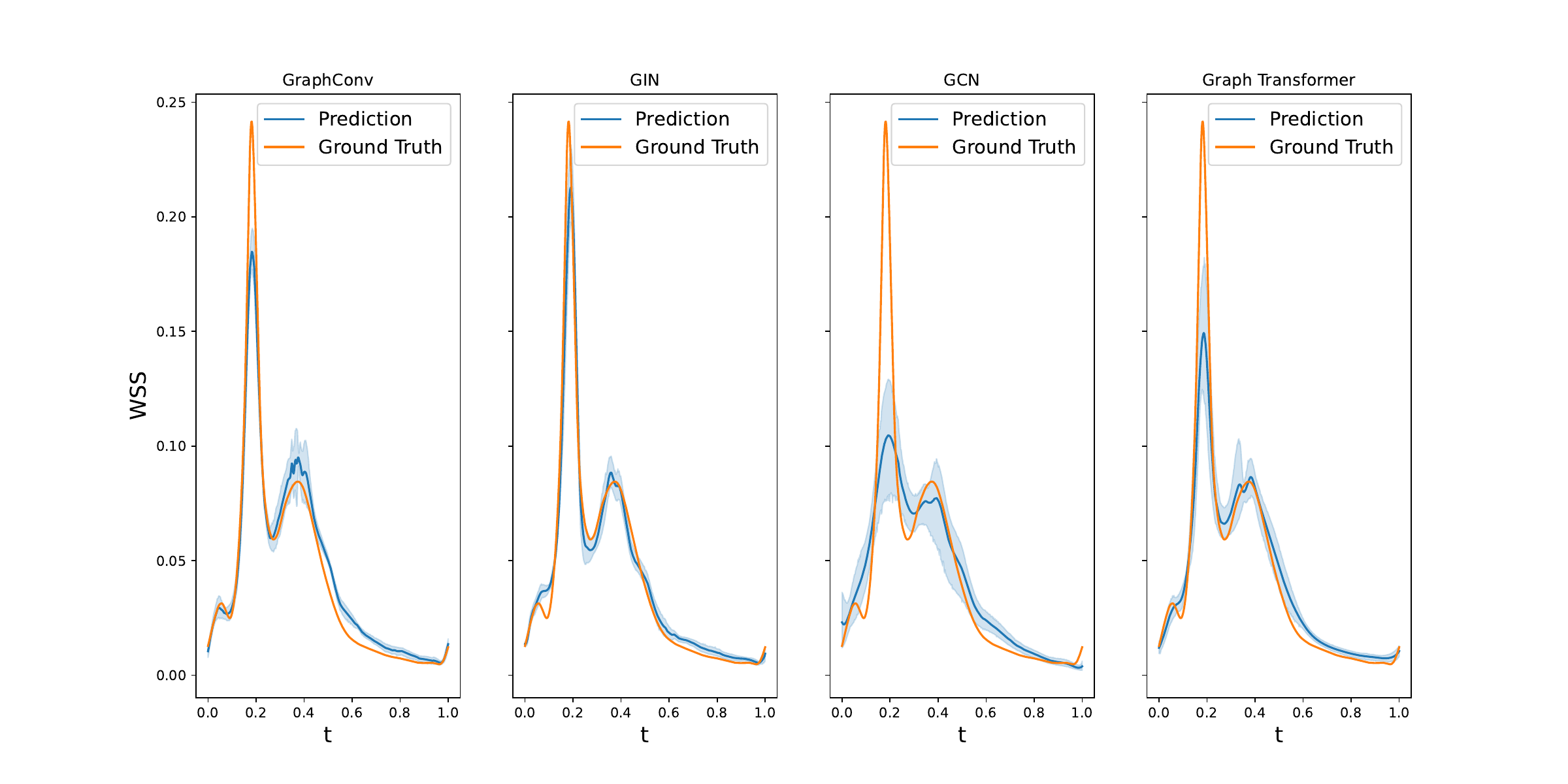}
    \caption{Prediction of the WSS index in time by different GNN modules on node 1415.}
    \label{fig:node_1415}
\end{figure}

From a qualitative point of view, we compare how GNNs with different modules perform on a node-wise prediction. Nodes were selected all across the surface mesh, to provide a comprehensive overview of the prediction capabilities on different zones of the aneurysm wall. Results correspondent to selected nodes all over the aneurysm surface mesh are reported in Figures \ref{fig:node_60}-\ref{fig:node_1415}. 

The selected node-wise predictions highlight how Graph Transformers better catch the peaks and the overall shape of the curve, compared to the other GNN modules. It is worth to notice that GNN with GraphConv modules are able to well predict the periodicity of the boundary conditions, but we will see in next paragraph that also GNNs with Graph Transformers modules possess this property, as long as a greater number of layers is provided.

Based on our preliminary analysis, from now on the implemented GNNs will make use of Graph Transformers modules.

\begin{figure}[!ht]
    \centering \includegraphics[width = \linewidth]{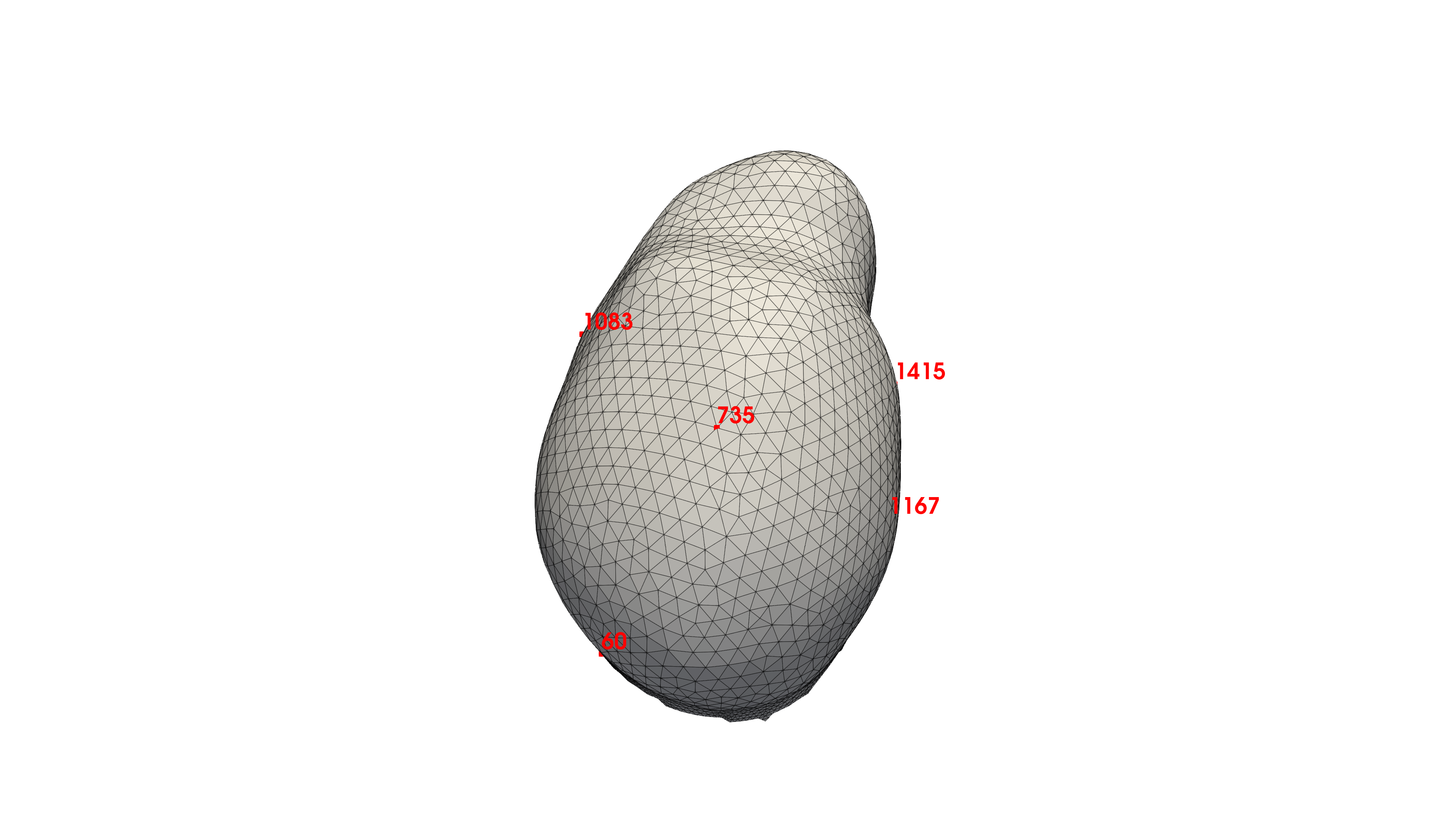}
    \caption{Selected nodes on which the GNN architecture is qualitatively assessed, comparing the prediction of the WSS index by different GNN modules.}
    \label{fig:selected}
\end{figure}

\newpage
 
\subsection{Ablation study on the number of layers}
Once the GNN module to employ has been established, 
we perform the task letting the \textit{number of layers} of the GNN model vary in the set $\{2,3,4,5\}$ while keeping the hidden dimension fixed. As theoretically demonstrated in literature, the expressive power of the Graph Neural Network may ideally rely solely on the number of message passing iterations (namely, layers) that we define for our model \cite{d2024approximation} \cite{azizian2020expressive} \cite{loukas2019graph}, while generalization capabilities could be affected by too large values of the hidden dimension \cite{d2024vc}; nevertheless, we acknowledge that an optimal choice of this hyperparameter could improve the performances; this could be a subject for further investigations in future works. For computational limitations of our hardware, we keep the hidden dimension fixed to $\mathsf{hd}=32$.

We compute statistics on $\text{err}_{\text{TAWSS}}$ to select the best choice of the number of layers. Statistics are reported in Table \ref{tab:TAWSS_layers_task1}.

\begin{table}[!ht]
    \centering
        \caption{Main statistics of $\text{err}_{\text{TAWSS}}$ over 5 runs on Task 1, for different numbers of employed layers. }
    \begin{tabular}{|c||c|c|c|c|}
    \hline
         & \textbf{mean} & \textbf{std} & \textbf{min} & \textbf{max} \\
       \hline
        \textbf{2 layers} & 0.364210 & 0.050328 & 0.330999 & 0.453135 \\
        \textbf{3 layers} & 0.422534 & 0.042029 & 0.352242 & 0.456900 \\
        \textbf{4 layers} & 0.323707 &  0.064259 & 0.259841 & 0.430129 \\
        \textbf{5 layers}  & 0.403281  & 0.068852 & 0.340992 & 0.504480
        \\ \hline
    \end{tabular}
    \label{tab:TAWSS_layers_task1}
\end{table}

\begin{figure}[!ht]
\centering\includegraphics[width=\linewidth]{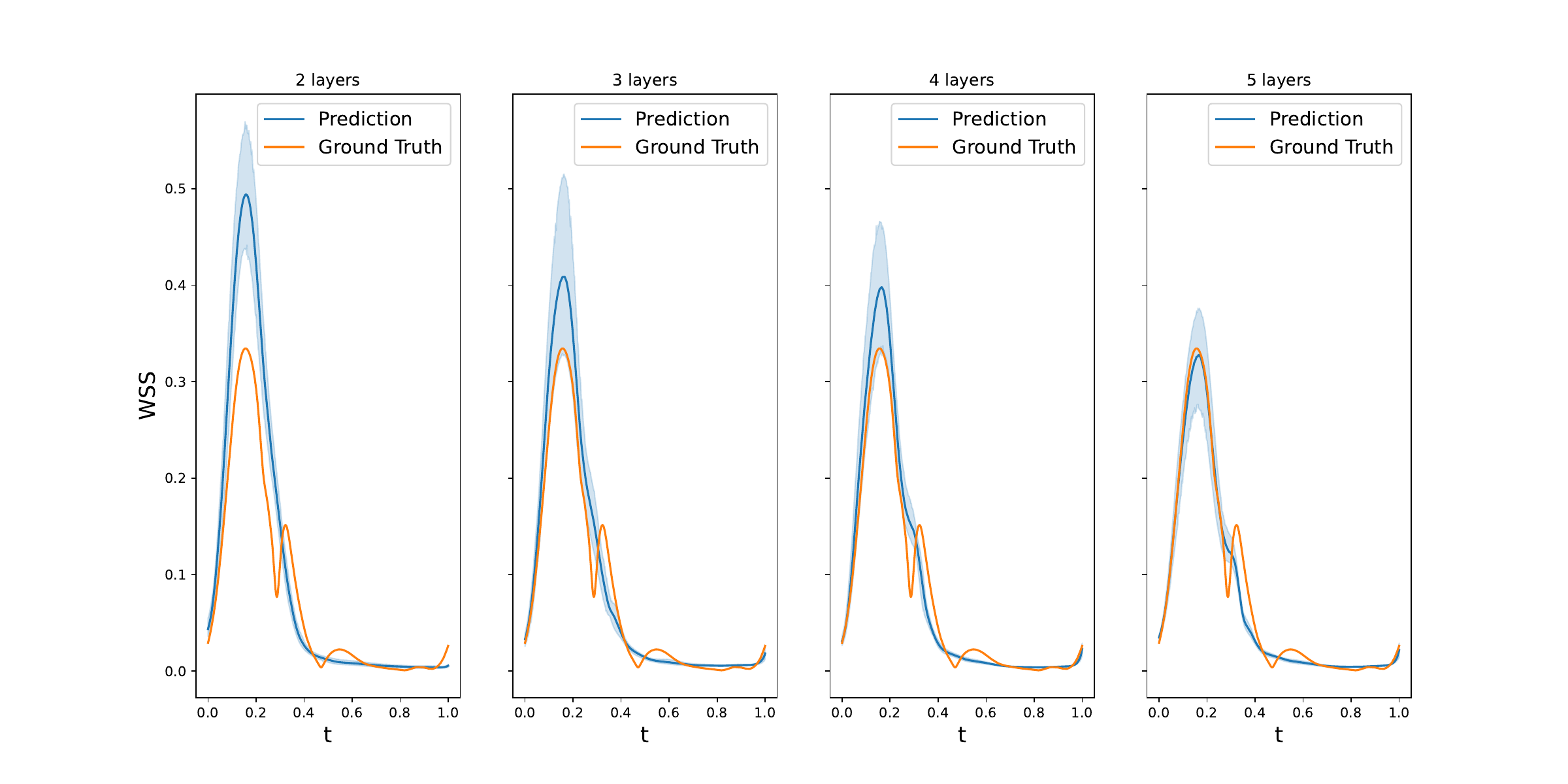}
    \caption{Prediction of the WSS index in time by GNNs with different number of layers on node 60.}
    \label{fig:node_60_l}
\end{figure}

\begin{figure}[!ht]
\centering\includegraphics[width=\linewidth]{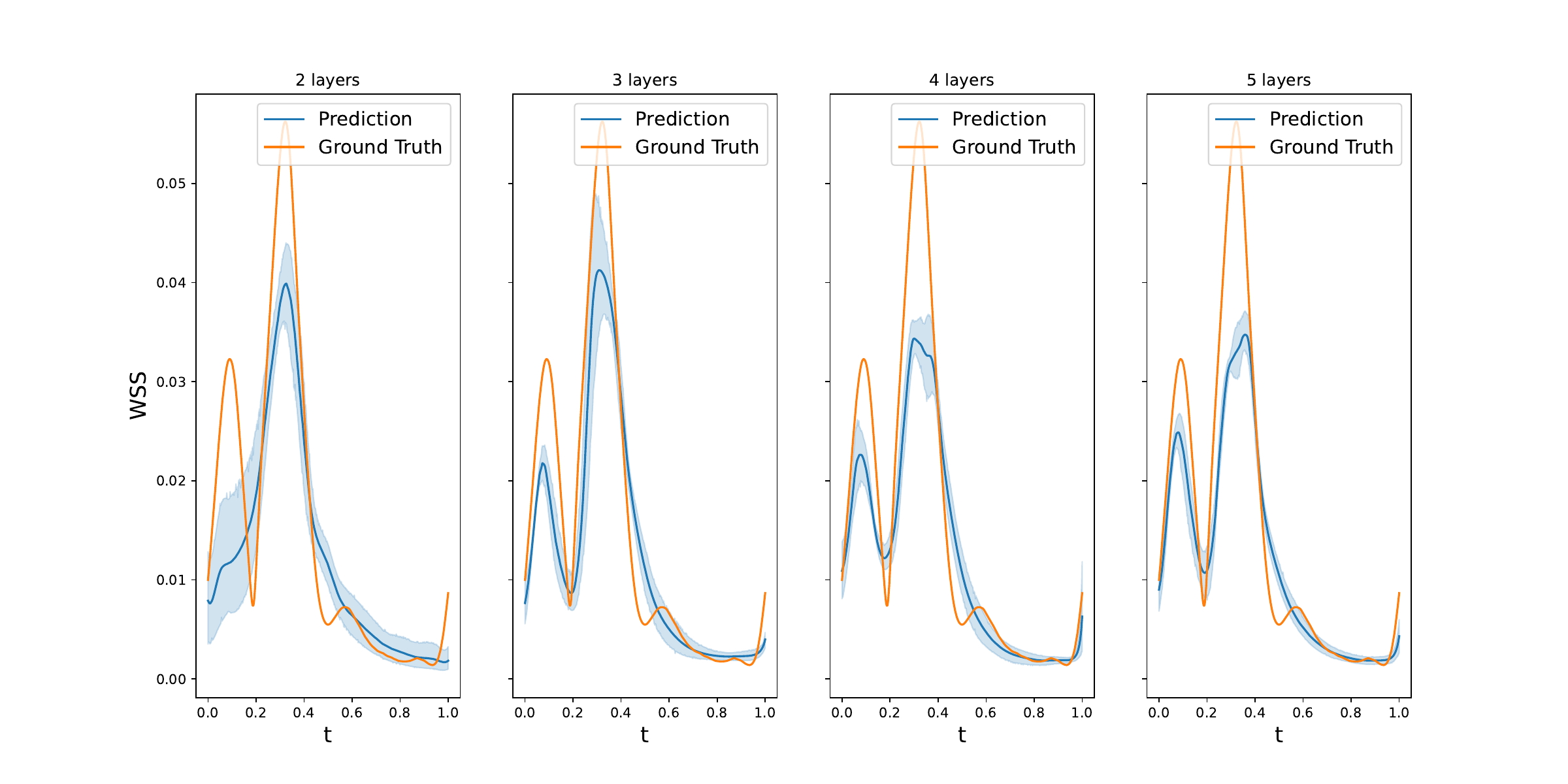}
    \caption{Prediction of the WSS index in time by GNNs with different number of layers on node 735.}
    \label{fig:node_735_l}
\end{figure}

\begin{figure}[!ht]
\centering\includegraphics[width=\linewidth]{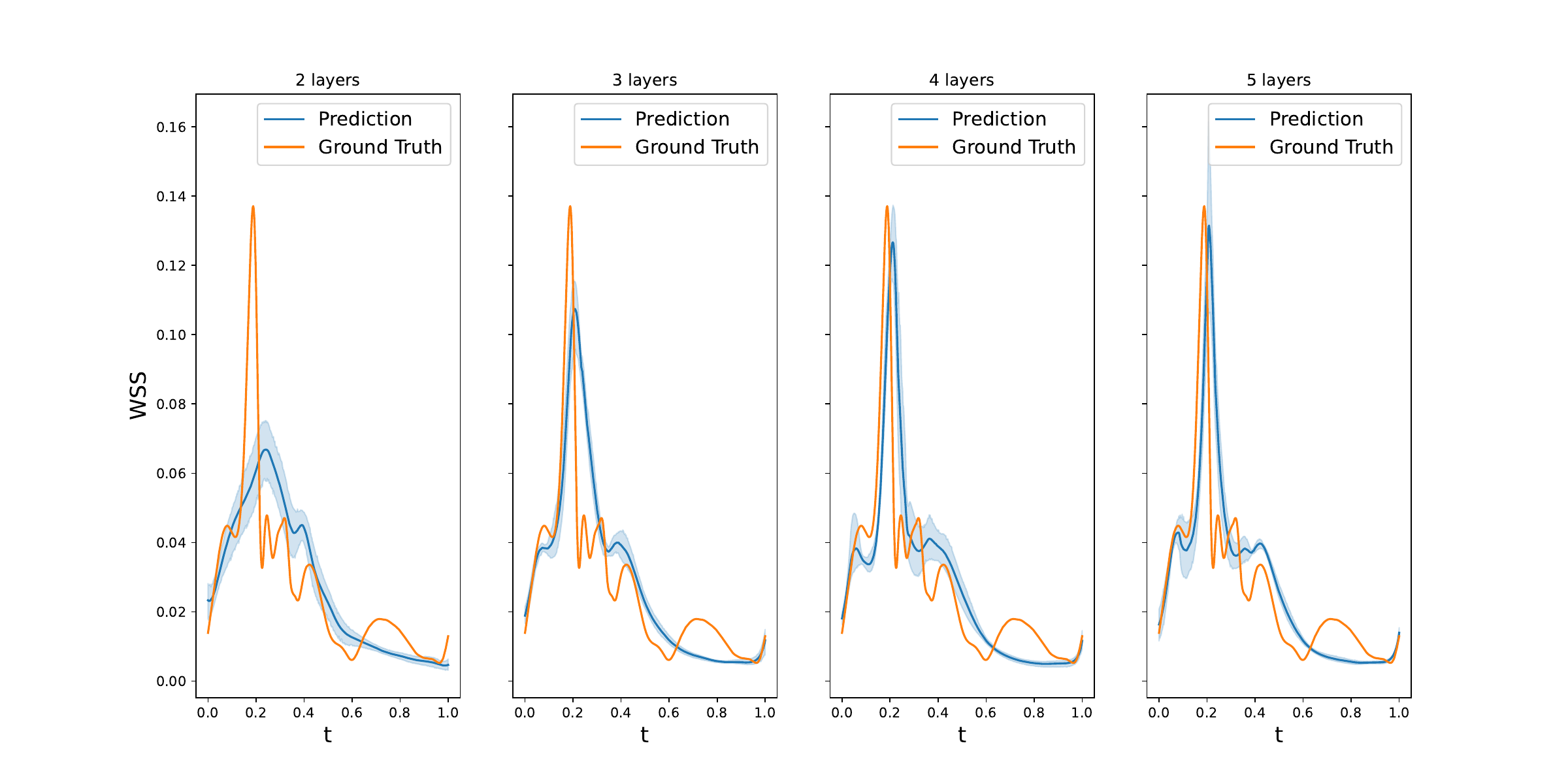}
    \caption{Prediction of the WSS index in time by GNNs with different number of layers on node 1083.}
    \label{fig:node_1083_l}
\end{figure}

\begin{figure}[!ht]
\centering\includegraphics[width=\linewidth]{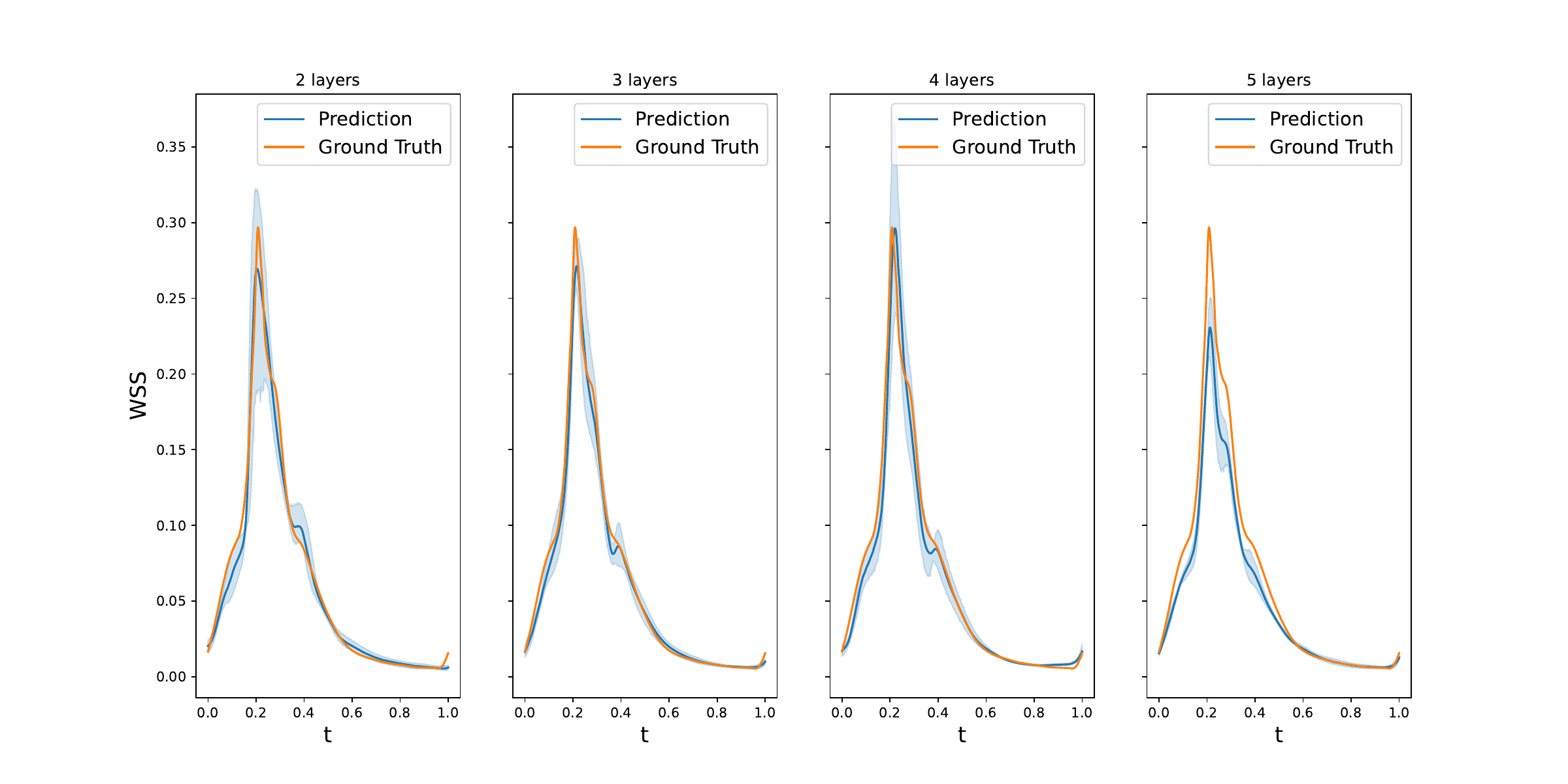}
    \caption{Prediction of the WSS index in time by GNNs with different number of layers on node 1167.}
    \label{fig:node_1167_l}
\end{figure}

\begin{figure}[!ht]
\centering\includegraphics[width=\linewidth]{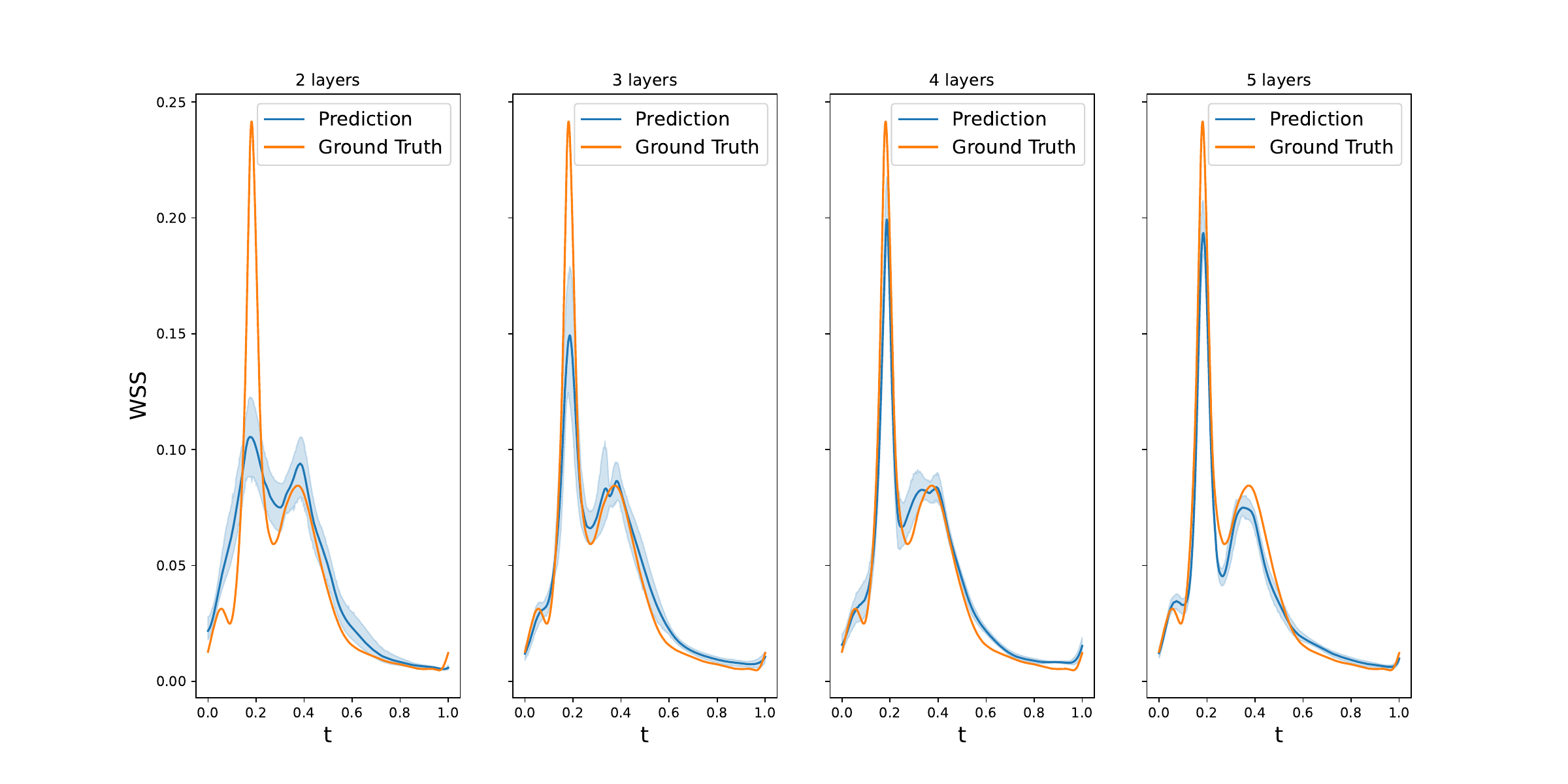}
    \caption{Prediction of the WSS index in time by GNNs with different number of layers on node 1415.}
    \label{fig:node_1415_l}
\end{figure}

The lowest mean value of $\text{err}_{\text{TAWSS}}$ is achieved with a GNN made of 4 layers. The (limited) node-wise qualitative comparison suggests that a greater number of layers helps in better learning the peak distribution and magnitude (Figures \ref{fig:node_60_l} - \ref{fig:node_1415_l})

 \newpage

 \newpage

\subsection{Interpolation task: numerical results}

\begin{figure}[!ht]
    \centering
    \includegraphics[width=\linewidth]{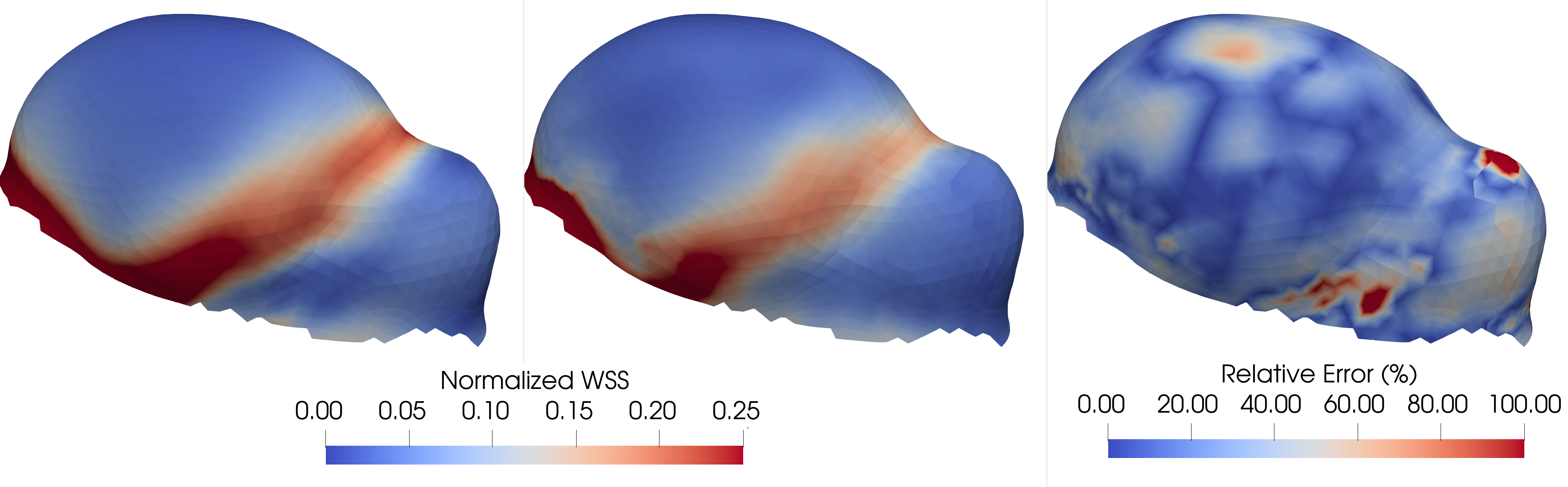}
    \caption{Ground truth (left) vs prediction (center) vs relative error percentage (left) on WSS index over a TAA of 75\% size percentage, at systolic time. The prediction is made by a GNN with 4 layers. In the figure, the WSS has been rescaled in $[0,0.25]$ to make the comparison clearer.}
    \label{fig:WSS_ground_vs_pred}
\end{figure}

The WSS prediction on the 75\% size TAA by the best GNN with 4 layers is shown in Figure \ref{fig:WSS_ground_vs_pred} in comparison with the ground truth, at timestamp $t=0.186$, which corresponds to the systolic peak. 

\begin{figure}[!ht]
    \centering
    \includegraphics[width=\linewidth]{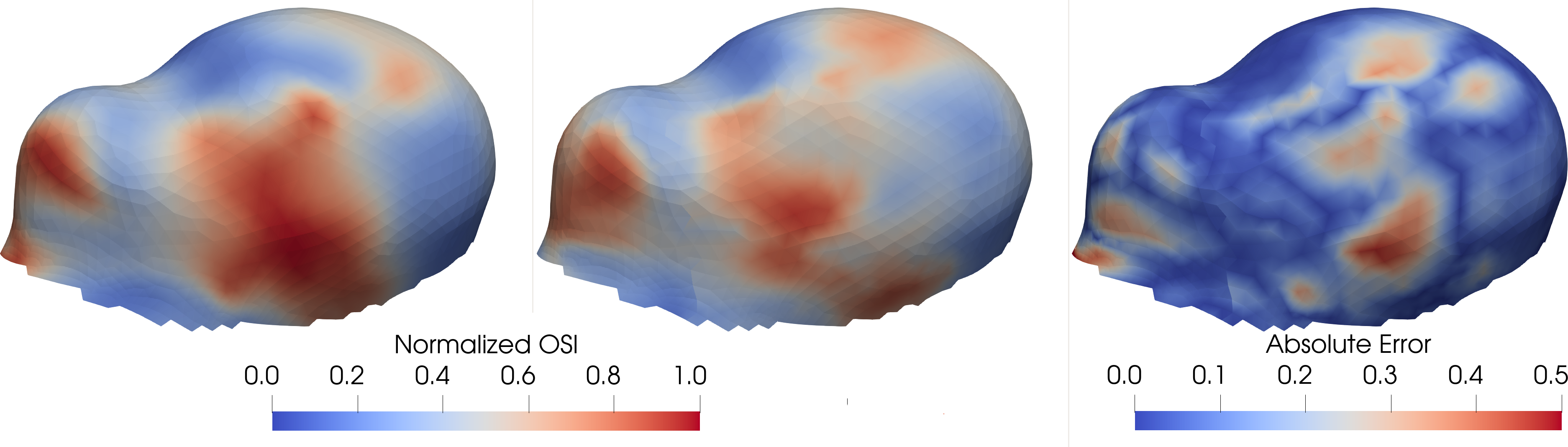}
    \caption{Ground truth (left) vs prediction (center) vs absolute error (left) on OSI index over a TAA of 75\% size percentage. The prediction is made by a GNN with 4 layers.}
    \label{fig:OSI_ground_vs_pred}
\end{figure}

\begin{figure}[!ht]
    \centering
    \includegraphics[width=\linewidth]{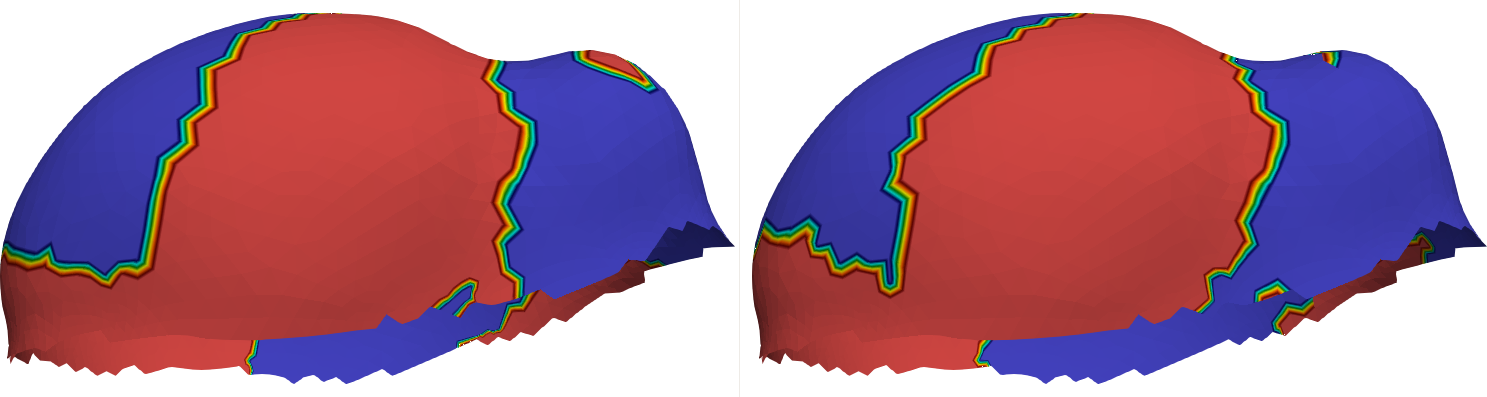}
    \caption{OSI at threshold  0.3 over a TAA of 75\% size percentage. Ground truth (left) vs prediction (right).}
    \label{fig:OSI_criterion}
\end{figure}

The prediction on the normalized OSI at the final timestamp (as defined in Equation \ref{eq:OSI}) by the same architecture is also provided  (Figure \ref{fig:OSI_ground_vs_pred}). Additionally, a threshold $\mathsf{thresh} = 0.3$ is applied to the normalized OSI, to highlight the accuracy of the prediction in high-risk areas (Figure \ref{fig:OSI_criterion}) \cite{xu2020improvement}. 

The GNN is able to well predict qualitatively the areas where the peaks of the stress indices will be reached in the systolic phase. The good performance of the predictor is additionally highlighted by the node-wise predictions in selected nodes, where the overall trend is well caught modulo the small oscillations.

\subsection{Extrapolation task: numerical results}
\begin{figure}[!ht]
    \centering
    \includegraphics[width=\linewidth]{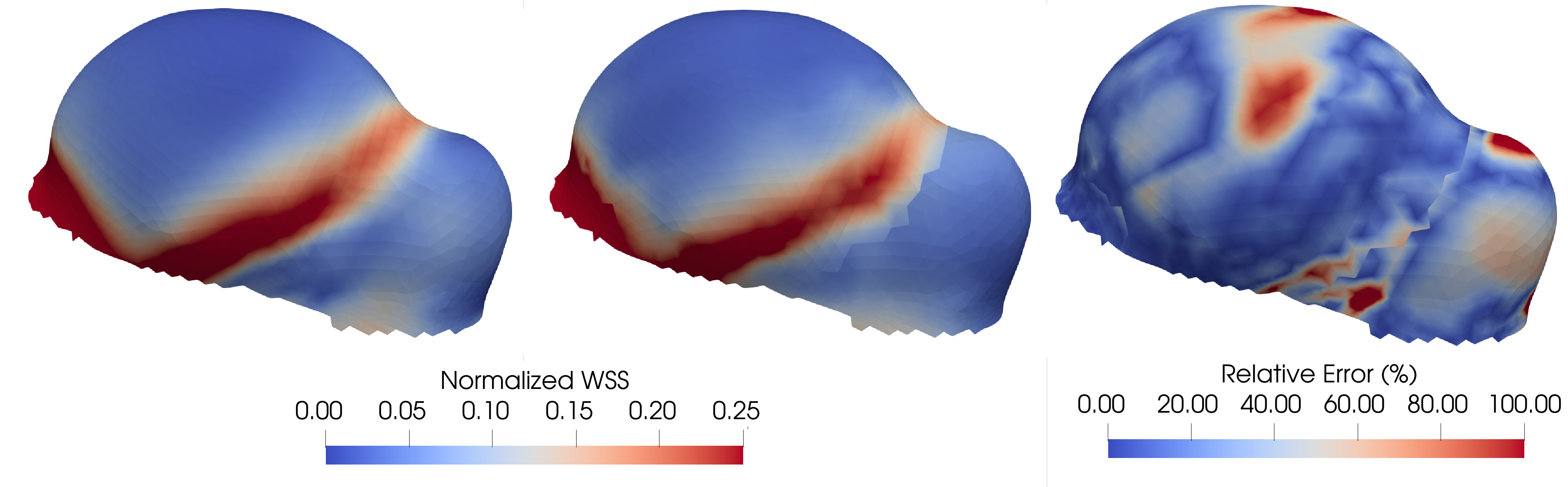}
    \caption{Ground truth (left) vs prediction (center) vs relative error percentage (left) on WSS index over a TAA of 100\% size percentage, at systolic time. The prediction is made by a GNN with 4 layers. In the figure, the WSS has been rescaled in $[0,0.25]$ to make the comparison clearer.}
    \label{fig:WSS_ground_vs_pred_extra}
\end{figure}

\begin{figure}[!ht]
    \centering
    \includegraphics[width=\linewidth]{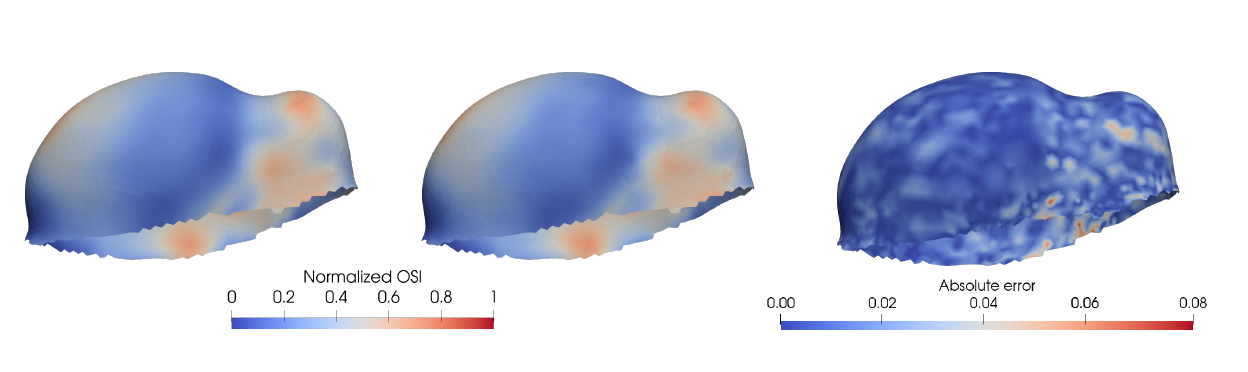}
    \caption{Ground truth (left) vs prediction (center) vs absolute error (left) on OSI index over a TAA of 100\% size percentage. The prediction is made by a GNN with 4 layers.}
    \label{fig:OSI_ground_vs_pred_extrap}
\end{figure}

\begin{figure}[!ht]
    \centering
    \includegraphics[width=\linewidth]{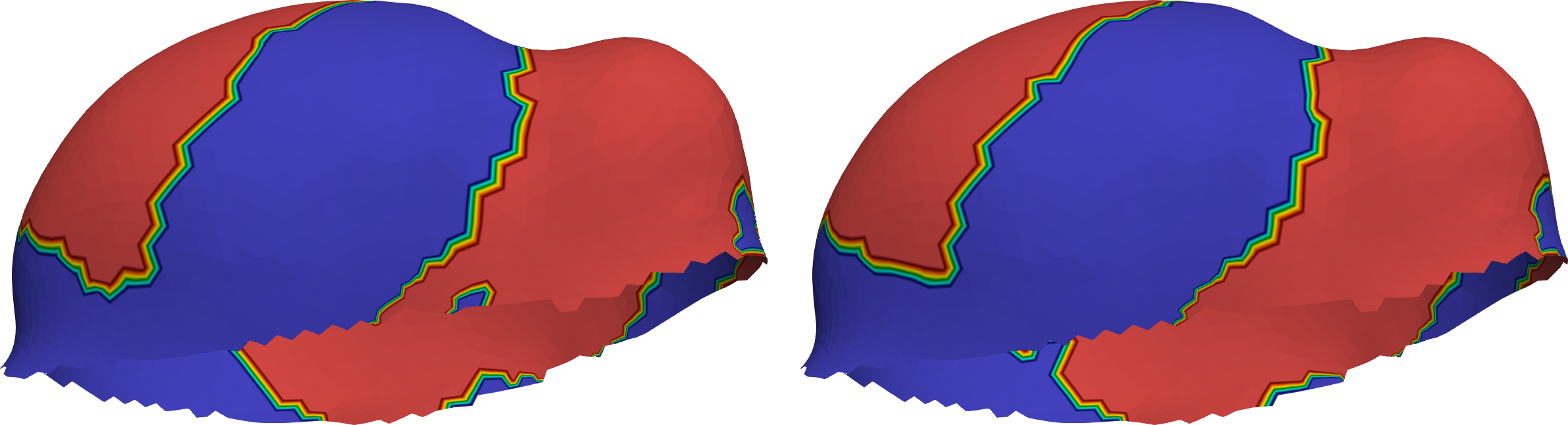}
    \caption{OSI at threshold  0.3 over a TAA of 100\% size percentage. Ground truth (left) vs prediction (right).}
    \label{fig:OSI_criterion_extrap}
\end{figure}

For the extrapolation task we show the prediction on WSS (at systolic peak) and OSI  (at the final timestamp) with and without thresholds (Figures \ref{fig:WSS_ground_vs_pred_extra}, \ref{fig:OSI_ground_vs_pred_extrap} and \ref{fig:OSI_criterion_extrap}, respectively). Both indices are normalized in $[0,1]$.

\begin{figure}[!ht]
    \centering
    \includegraphics[width=1\linewidth]{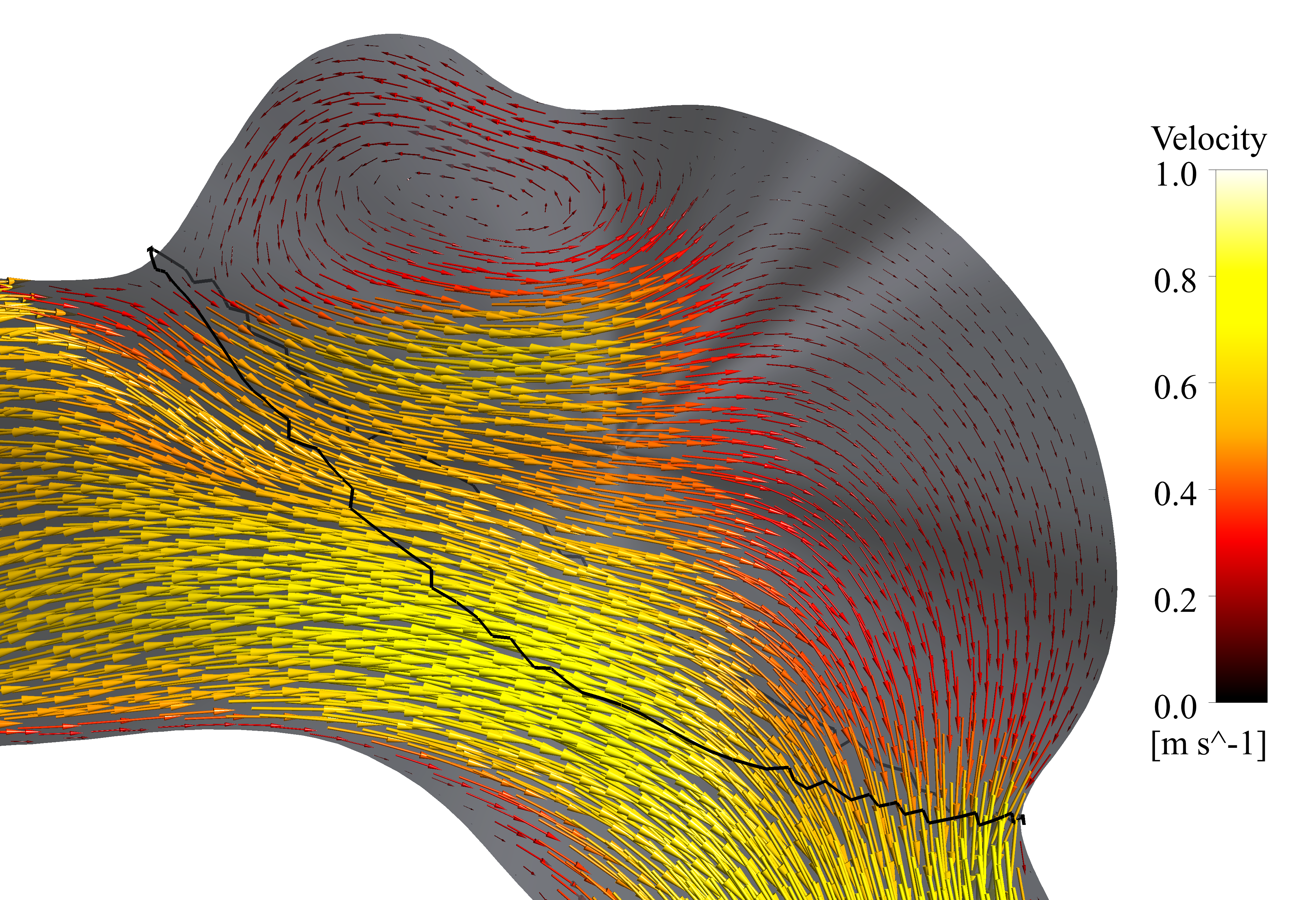}
    \caption{Velocity vectors inside aneurysm at the systolic peak.}
    \label{fig:VectorAtSystolic}
\end{figure}

Similar trends in the error distribution of WSS and OSI are observed in the extrapolation case, similar to those in interpolation. The network effectively predicts the regions of high concentration of WSS (Figure \ref{fig:WSS_ground_vs_pred_extra}) and OSI (Figure \ref{fig:OSI_ground_vs_pred_extrap}). The range of relative errors for WSS in both interpolation and extrapolation cases is comparable, with notably higher absolute errors in the aneurysm bleb region. This can be attributed to the complex flow dynamics in that region compared to the aneurysm sac. Specifically, Figure \ref{fig:VectorAtSystolic} illustrates the velocity vectors at the systolic peak within the aneurysm, highlighting more pronounced vortical structures in the bleb and greater flow uniformity in the sac. This complexity amplifies flow non-linearity, thereby influencing predictions and resulting in higher error values.

While the OSI prediction error distribution is similar for interpolation and extrapolation, a more detailed quantitative analysis reveals that the maximum absolute error in the extrapolation case is 0.08—significantly lower than the interpolation case, which has a maximum error of 0.5, as shown in Figure \ref{fig:OSI_ground_vs_pred}. The error remains concentrated around the aneurysm bleb, again linked to the heightened non-linearity of the flow field in this region. As previously noted, clinical guidelines identify an OSI value exceeding 0.3 as indicative of rupture risk. Based on this criterion, Figure \ref{fig:OSI_criterion_extrap} compares the FOM and GNN results, demonstrating a high level of agreement.

\section{Conclusion}
\label{sec:conclusion}

In this paper, we developed a novel framework for the prediction of shear stress indices, such as the WSS, integrating Full Order Models techniques with surrogate models; in particular, we obtained high fidelity data from blood flow modeling via Finite Volumes discretization in order to train Graph Neural Networks. This approach overcomes the traditional mesh refinement problem, as GNNs can yield instantaneous evaluations on the input meshes, which can be made of different numbers of nodes. GNNs exhibit good qualitative performance in predicting WSS and OSI, both on interpolation and extrapolation tasks, providing an alternative tool to monitor shear stress peaks at different volume sizes and consequently to quantify the risk of ruptures or dissections in the TAA. 

Our method is not free from open issues. Firstly, the implemented GNNs are permutation equivariant, but not SE3 equivariant. Although our results do not show limitations arising from the lack of this property, the implementation of SE3 equivariance would make our approach more robust to data assimilation (see, for instance, the approach used in \citep{suk2024mesh}). As our model leverages space coordinates as relevant features for an effective prediction, same data taken in a different system of coordinates could be subject to a bad predictive behaviour. A SE3 equivariant model would inherently possess the desirable feature of being equivariant to rototranslations in space. \\
The individual performances of the nodes need to be enhanced to ensure accurate local predictions. One way to achieve this is by treating the data as temporal sequences and using temporal models, namely, Temporal GNNs \cite{longa2023graph}. This approach would involve a significant preprocessing stage, where we'd need to define a sliding window of a certain length to create temporal graph sequences that are long enough to capture temporal behavior, yet short enough to be processed by a relatively simple Temporal GNN, in order to avoid issues like vanishing gradients \cite{taheri2019predictive}. Another option would be to process the high-resolution data as point clouds, allowing us to apply state-of-the-art methods for prediction tasks on point clouds, such as PointNets \cite{qi2017pointnet}.
Eventually, with more advanced hardware sources, we could fine-tune our model via grid-search methods to find the optimal configuration. Another notable limitation of this study is the reliance on data from a single patient. While the results provide valuable insights, they may not fully capture the variability across a broader population. Future work should focus on validating the findings using data from multiple patients to enhance the generalizability of the proposed approach. Expanding the dataset will also allow for a more robust assessment of inter-patient variability and its implications on the model's predictive performance.

\section*{Acknowledgements}
G.A.D., P.C.A., and G.R. acknowledge the support provided by the European Union - NextGenerationEU, in the framework of the iNEST - Interconnected Nord-Est Innovation Ecosystem (iNEST ECS00000043 – CUP G93C22000610007) project and its CC5 Young Researchers initiative. S.S. and G.R. acknowledge the support provided by PRIN "FaReX - Full and Reduced order modelling of coupled systems: focus on non-matching methods and automatic learning" project, and by INdAM-GNCS 2019–2020 projects and PON "Research and Innovation on Green related issues" FSE REACT-EU 2021 project. The views and opinions expressed are solely those of the authors and do not necessarily reflect those of the European Union, nor can the European Union be held responsible for them. P.C.A. also acknowledges the INdAM-GNCS Project ``Metodi numerici efficienti per problemi accoppiati in sistemi complessi'' (CUP E53C24001950001). In addition, the authors would like to acknowledge INdAM–GNCS.

\section*{Conflict of interest}
The authors declare no potential conflict of interests.

\section*{Ethics approval and consent to participate}
Not applicable.

\section*{Consent for publication}
Not applicable.

\section*{Data availability}
Patient-specific raw data are not publicly available.

\section*{Materials availability}
Not applicable.

\section*{Code availability}
The code implementing GNNs is available at \url{https://github.com/AleDinve/aneurysm_gnn/}.

\section*{Author contributions}
GAD: Conceptualization, Methodology, Software, Formal analysis, Investigation, Data Curation, Writing - Original Draft. SM: Conceptualization, Methodology, Software, Formal analysis, Writing - Original Draft. SS: Conceptualization, Methodology, Formal analysis, Writing - Original Draft. PCA: Conceptualization, Methodology, Formal analysis, Writing - Original Draft, Supervision. GR: Writing - Review \& Editing, Project management, Funding acquisition.

\appendix
\section{GNN modules}\label{app:gnn_modules}
\begin{itemize}
    \item The GraphConv module \cite{hamilton2017representation} \cite{morris2019weisfeiler}, which is the simplest GNN module proven to match the expressive power of the 1-WL test. The updating scheme is designed as follows:
    \begin{equation}\label{eq:gconv}
        \mathbf{h}_v^{(t+1)} = f \big ( \mathbf{W}_{\mathsf{comb}} \mathbf{h}_v^{(t)} + \mathbf{W}_{\mathsf{agg}} \sum \limits_{u \in \mathsf{ne}(v)} \mathbf{h}_u^{(t)}  + \mathbf{b} \big )
    \end{equation}
    \item The Graph Convolutional Network (GCN) module \cite{kipf2016semi} is maybe the most famous GNN framework. Hidden features are updated by means of convolutions with the adjacency matrix of the graph and its vertex features:
    \begin{equation}\label{eq:gcn_full}
        \mathbf{H}^{(t+1)} = \hat{\mathbf{D}}^{-1/2} \hat{\mathbf{A}} \hat{\mathbf{D}}^{-1/2} \mathbf{H} \mathbf{W}
    \end{equation}
    where $\hat{\mathbf{A}} = \mathbf{A} + \mathbf{I}$ denotes the adjacency matrix with inserted self-loops and $\hat{D}_{ii} = \sum \limits_{j=0}^N \hat{A}_{ij}$ the entries of its diagonal degree matrix. 
    The node-wise formulation of a GCN module is given by:
    \begin{equation}\label{eq_gcn_node}
        \mathbf{h}_v^{(t+1)} = \mathbf{W}^T \sum \limits_{u \in \mathsf{ne}(v) \cup \{v\}} \frac{1}{\sqrt{\hat{d}_u \hat{d}_v}}\mathbf{h}_u^{(t)}
    \end{equation}
    \item The Graph Isomorphism Network module (GIN) \cite{xu2018powerful} has been the first GNN module with more expressive power than the 1-WL test. It applies an MLP to a linear combination of the node hidden features:
    \begin{equation}\label{eq:gin}
        \mathbf{h}_v^{(t+1)} = \mathsf{MLP} \big ( (1+\epsilon) \mathbf{h}_v^{(t)} + \sum \limits_{u \in \mathsf{ne}(v)} \mathbf{h}_u^{(t)} \big )
    \end{equation}
    \item The Graph Transformer module \cite{shi2020masked} integrates the multi-head attention mechanism of Transformers \cite{vaswani2017attention} into graph learning. The scheme formulation refines the one in Equation \ref{eq:gconv} adding a weight determined by the multi-head:
    \begin{equation}\label{eq:graph_transformer}
        \mathbf{h}_v^{(t+1)} = f \big ( \mathbf{W}_{\mathsf{comb}} \mathbf{h}_v^{(t)} + \mathbf{W}_{\mathsf{agg}} \sum \limits_{u \in \mathsf{ne}(v)} \alpha_{uv} \mathbf{h}_u^{(t)}  + \mathbf{b} \big )
    \end{equation}
    where the attention coefficients $\alpha_{uv}$ are computed via multi-head dot product attention:
    \begin{equation}\label{eq:multihead_attention}
        \alpha_{uv} = \mathsf{softmax} \left( \frac{(\mathbf{W}_{\mathsf{mh}1}\mathbf{h}_v^{(t)})^T (\mathbf{W}_{\mathsf{mh}2}\mathbf{h}_u^{(t)}) }{\sqrt{d}} \right) 
    \end{equation}
\end{itemize}

\bibliographystyle{elsarticle-num}
\bibliography{references}

\end{document}